\documentclass[aps,prb,twocolumn,floatfix, 10pt]{revtex4-1}

\usepackage{graphicx, bm, amsmath, amsfonts,amssymb,amsthm}

\usepackage{float}
\usepackage{color}
\usepackage{ulem}

\newcommand{\Z}{\mathbb{Z}}

\newcommand{\mfg}{{\mathfrak{g}}}

\newcommand{\tomega}{{\tilde{\omega}}}
\renewcommand{\d}{\partial}

\newcommand{\beq}{\begin{equation}}
\newcommand{\eeq}{\end{equation}}
\newcommand{\ba}{\begin{array}{ccc}}
\newcommand{\ea}{\end{array}}
\newcommand{\bea}{\begin{eqnarray}}
\newcommand{\eea}{\end{eqnarray}}

\newcommand{\mz}{\mathbb{Z}}

\newcommand{\bfg}{{\bf{g}}}
\newcommand{\bfh}{{\bf{h}}}
\newcommand{\bfk}{{\bf{k}}}
\newcommand{\bfl}{{\bf{l}}}

\graphicspath{{figures/}}

\begin{document}

\title{Surface Topological Order and a new 't Hooft Anomaly of Interaction Enabled 3+1D Fermion SPTs}

\author{Lukasz Fidkowski}
\affiliation{Department of Physics, University of Washington, Seattle WA  98195, USA}

\author{Ashvin Vishwanath}
\affiliation{Department of Physics, Harvard University, Cambridge MA 02138, USA}

\author{Max A. Metlitski}
\affiliation{Department of Physics, Massachusetts Institute of Technology, Cambridge, MA  02139, USA}

\begin{abstract}

Symmetry protected topological (SPT) phases are well understood in the context of free fermions and in the context of interacting but essentially bosonic models. Recently it has been realized that intrinsically fermionic  SPTs exist which only appear in interacting models. Here we show that the 3+1 dimensional realizations of these phases {have surface states} characterized by a new 't Hooft anomaly, captured by a $H^3(G, \mz_2)$ class. This is encoded in the anomalous action of symmetry on the surface states with topological order, which must necessarily permute the anyons. We discuss in detail an example with symmetry group $G = \mz_2 \times \mz_4$. Using a network model of the surface we  derive a candidate surface topological order given by a $\mz_4$ gauge theory. We relate our findings to anomalies valued in $H^3$ with various coefficients introduced previously in both bosonic and fermionic settings, and describe a general framework that unifies these various anomalies.
\end{abstract}

\maketitle

\section{Introduction.}

In the last few years we have learned a lot about possible realization of symmetry in quantum systems. Symmetry protected topological phases (SPTs) have played an important role in this progress \cite{Chen2d, DW, Levin_Gu}. These are gapped states of matter protected by a symmetry $G$: once the symmetry is broken, one can continuously connect an SPT phase to a trivial product state. Much of the interesting physics of SPT phases comes from their boundary, which must necessarily be non-trivial; the boundaries of 1+1 and 2+1 dimensional SPTs are  either gapless or symmetry broken,\cite{CZX} while in 3+1 and higher dimensions a new possibility exists: the boundary may be gapped and symmetry preserving at the cost of supporting intrinsic topological order.\cite{Vishwanath2013, Metlitski2013, Chen2013a, Bonderson2013, Wang2013a, Fidkowski2013, Burnell2013, ProjS} Moreover, in all cases the boundary of an SPT state is ``anomalous": the way it realizes the symmetry cannot be mimicked without the bulk. In fact, even though depending on the boundary dynamics different boundary phases may exist, all of these phases share a common anomaly that must match the bulk. For SPT phases of bosons a  systematic understanding of this matching exists in several cases: i) projective symmetry on the boundary of a 1+1D SPT \cite{Chen1d, Fidkowski1d, TurnerBerg} ii) a CFT on the boundary of a 2+1D SPT \cite{LuV, Bultinck, Yin}; iii) symmetric intrinsic topological order on the boundary of a 3+1D SPT  - this is the case which will be of interest to us here.\cite{ProjS, WLL}

How does one characterize a 2+1D intrinsic topological order in the presence of a global symmetry (also known as symmetry enriched topological order (SET))?  In fact, a systematic algebraic theory of 2+1D SETs has recently been developed \cite{Maissam2014, Teo_Hughes, Tarantino2015}. The following data goes into this theory: i) the anyon content  and the braiding and fusion rules; ii) how the anyons are permuted under the action of the symmetry; iii) fractional quantum numbers carried by the anyons. It turns out that not all realizations of symmetry, which are seemingly consistent with anyon fusion and braiding can exist in a strictly 2+1D system. In particular, some assignments of anyon fractional quantum numbers cannot be realized in 2+1D \cite{ProjS}. The anomaly for a given symmetry fractionalization is given by an element  $\mu \in H^4(G, U(1))$, where $G$ is the symmetry group and $H^D(G, U(1))$ is the $D$-dimensional cohomology group with $U(1)$ coefficients \cite{ENO}. (Here and below, unless otherwise noted, we specialize to the case where $G$ is a unitary, discrete, internal symmetry.)  A very physical interpretation of this anomaly exists: 3+1D SPT phases of bosons are also classified by $H^4(G, U(1))$, and it is believed that a 2+1D SET with anomaly $\mu$ can live on the surface of the corresponding 3+1D SPT. This belief is supported by a large class of examples.\cite{ProjS, WLL}

In the discussion above we have glossed over the fact that an even more severe anomaly of the symmetry $G$ may be present. Namely, the permutation of the anyons under the action of the symmetry might itself be anomalous, even though it preserves the anyon fusion and braiding. 
 This anomaly is measured by an element $O \in H^3(G, A)$, where $A$ is the group of Abelian anyons in the topological order. Such anomalous 2+1D SETs cannot exist at the boundary of a 3+1D SPT, but may be interpreted in a certain way as living on the boundary of a 3+1D SET \cite{Fidkowski_Vishwanath} (with the caveat that certain anyons are confined to endpoints of 3+1D loop-like excitations), or a 3+1D phase with higher form symmetry.\cite{CurtRyan, Kapustin_higher_form, H32group}
 
One may ask, what is the corresponding situation for phases of fermions. While non-interacting SPTs of fermions to which the famous conventional topological insulators belong have been understood for some time now\cite{KaneTIReview, KitaevNI,LudwigNI} and have in many ways precipitated the  study of SPTs, the general  classification of interacting fermion SPTs is fairly recent. The case of 3+1D fermion SPTs is particularly interesting since here for a unitary internal symmetry in the absence of interactions no non-trivial phases exist. 
Currently, general interacting fermion SPTs in 3+1D with symmetry $G \times \Z^f_2$ are believed to be classified by three inputs \cite{Kapustin2017, Wang_Gu, CTW}: a co-cycle $\sigma \in H^2(G, \Z_2)$ and co-chains $\rho \in C^3(G, \Z_2)$ and $\mu \in C^4(G, U(1))$ satisfying certain algebraic conditions and modulo equivalence relations. For fixed $\sigma$, $\rho$, solutions to these conditions differ by $\mu' - \mu \in H^4(G, U(1))$, which physically corresponds to stacking on a boson SPT phase. Thus, the intrinsically fermionic physics comes from the inputs $\sigma$ and $\rho$. The case  $\sigma = 0$ gives so-called super-cohomology phases \cite{supercohomology}: here, $\rho$ must be a co-cycle in $H^3(G, \Z_2)$.  The case when the total symmetry group does not factor as $G \times \Z^f_2$ is less well understood, although see Ref. \onlinecite{CTW} for an approach for Abelian $G$ via the classification of three loop braiding statistics.

What kind of surface topological order (STO) do 3+1D fermion SPTs admit? While many examples of such surface states have been constructed for topological insulators and superconductors protected by time-reversal symmetry (possibly in conjunction with other unitary symmetries),\cite{ Fidkowski2013, Metlitski2013, Chen2013a, Bonderson2013, Wang2013a} no examples for the case of purely unitary symmetries are known. In fact, since for purely unitary internal symmetries the bulk is necessarily strongly interacting, here we don't yet know any surface states, either gapless or topologically ordered.  
Our goal is to construct STOs for this case and to characterize the STO anomaly. We focus on bulk fermion SPTs in super-cohomology with symmetry group $G \times \Z^f_2$. We expect the corresponding surface topological order to possess a new anomaly characterized by a co-cycle $H^3(G, \Z_2)$. Indeed, we define such an anomaly for a general 2+1D fermion topological order. Again, the data that specifies the fermion SET is: i) the anyon content; ii) how the symmetry permutes the anyons; iii) the fractionalization of symmetry on the anyons. The new $H^3(G, \Z_2)$ anomaly is associated with an obstruction to extending the symmetry fractionalization to fermion parity fluxes after gauging the $\Z^f_2$ symmetry. We construct a simple example of such an anomalous SET for symmetry group $G  = \Z_2 \times \Z_4$: the topological order is $\Z_4 \times \{1, f\}$, i.e. a $\Z_4$ gauge theory together with the physical fermion $f$. The subgroup $\Z_4 \subset G$ acts on the anyons by a non-trivial permuation, and the anyons carry a particular fractionalization of $G$. We conjecture that this topological order can live at the surface of a 3+1D fermion SPT. Indeed, for this symmetry, modulo bosonic SPTs, there is a single intrinsically fermionic super-cohomology phase corresponding to a non-trivial co-cycle $\rho \in H^3(G, \Z_2)$.\cite{Kapustin2017,CTW,JuvenSpin} 
We conjecture that the $\Z_4$ gauge theory above is a STO of this intrinsically fermionic SPT. In particular, we find that the anomaly co-cycle $\rho \in H^3(G, \Z_2)$ extracted from the topological order matches the bulk co-cycle. Further evidence in favor of our conjecture comes from a closely related example: a fermion crystalline SPT with symmetry group $\Z_2 \times \Z \times \Z^f_2$. Here, $\Z_2$ is an internal symmetry and $\Z$ is lattice translation symmetry along a particular direction. Such an SPT can be constructed by stacking layers of 2+1D fermion SPT with $\Z_2 \times \Z^f_2$ symmetry. The latter are classified by an integer $\nu \in \Z_8$ \cite{Qi, Ryu_Zhang, Gu_Levin}. We focus on the 3+1D crystalline SPT constructed by stacking layers with $\nu =2$. As we will explain, this is a close cousin of the non-crystalline $\Z_2 \times \Z_4 \times \Z^f_2$ 3+1D super-cohomology SPT discussed above.  For the crystalline SPT example, we can explicitly construct the STO by gapping out the gapless modes associated with the edges of the 2+1D layers. In this way, we obtain the $\Z_4$ gauge theory described above.

Having defined the $H^3(G,\Z_2)$ anomaly for general 2+1D fermion SETs, we may ask, does this anomaly have a relation to the $H^3$ anomaly of bosonic SETs \cite{Fidkowski_Vishwanath, Maissam2014}? Clearly, the two anomalies have a somewhat different physical interpretation: in the fermionic case the topological order can live on the surface of an SPT, while in the bosonic case it cannot. Furthermore, in the fermionic case we need to specify the fractionalization of the symmetry on the anyons, while in the bosonic case the anomaly appears already at the level of anyon permutation by the symmetry - i.e. no symmetry fractionalization is consistent with the anyon permutation. Nevertheless, the two anomalies can be united into a common framework that we introduce. One imagines a topological order where the permutation of the anyons by the symmetry is specified. Furthermore, a consistent symmetry fractionalization is specified for a {\it subset} of the anyons (closed under fusion). One then asks whether symmetry fractionalization can be extended to the rest of the anyons. The general $H^3$ anomaly is the obstruction to doing so. For the case of the original $H^3$ anomaly for bosonic SETs, the subset of anyons for which fractionalization is specified is just the identity particle (which carries no fractionalization). For the case of  fermionic SETs, one may consider the modular extension of the topological order. The anyons of the original fermionic SET are the subset of the modular extension on which fractionalization is specified, the remaining anyons of the modular extension are the fermion parity fluxes, and one asks if fractionalization can be consistently defined on them. Other previously considered examples \cite{KapustinThorngrenZ3} also neatly fit into this general framework.

This paper is organized as follows. In section \ref{sec:bulk}, we review the properties of the 3+1D fermion SPT with ${\mathbb Z}_2 \times {\mathbb Z}_4 \times {\mathbb Z}^f_2$ symmetry and its cousin crystalline SPT with ${\mathbb Z}_2 \times {\mathbb Z}\times {\mathbb Z}^f_2$ symmetry. We then introduce our proposed STO for these two SPTs in section \ref{sec:STO} and give a physical argument why such an STO is anomalous. Section \ref{sec:stack} gives an explicit construction of the STO for the crystalline SPT. Section \ref{sec:H3f} gives a general definition of the anomaly class $[\rho] \in H^3(G, \Z_2)$ for 2+1D fermionic topological orders. In section \ref{sec:H3comp}, we apply this definition to our conjectured topological order for the ${\mathbb Z}_2 \times {\mathbb Z}_4 \times {\mathbb Z}^f_2$ SPT and show that the resulting surface anomaly class $[\rho]$ matches the bulk class. In section \ref{sec:H3gen}, we discuss the general framework for $H^3$ type anomalies in 2+1D topological orders. We conclude with some open questions in section \ref{sec:disc}.

\section{${\mathbb Z}_2 \times {\mathbb Z}_4 \times {\mathbb Z}^f_2$ 3+1D fermion SPT and its crystalline cousin.}
\label{sec:bulk}

In this section, we review some properties of the 3+1D fermion SPT with internal symmetry ${\mathbb Z}_2 \times {\mathbb Z}_4 \times {\mathbb Z}^f_2$ and explain how it is related to the crystalline fermion SPT with symmetry ${\mathbb Z}_2 \times {\mathbb Z}\times {\mathbb Z}^f_2$, where ${\mathbb Z}_2$ is an internal symmetry and ${\mathbb Z}$ is translation along a given direction.

A key role in our discussion will be played by 2+1D fermion SPTs with $\mz_2 \times \mz^f_2$ symmetry, so we review them first. Recall, these have a $\mz_8$ classification\cite{Qi, Ryu_Zhang, Gu_Levin}. The phase with $\nu = 1$ can be constructed by stacking a $p_x + i p_y$ superconductor with constituent fermions charged under the $\mz_2$ symmetry and a $p_x - i p_y$ superconductor with constituent fermions neutral under $\mz_2$. The edge of the $\nu  =1$ phase then consists of a right-moving Majorana fermion ($c  =1/2$) charged under $\mz_2$ and a left-moving Majorana fermion neutral under $\mz_2$. Likewise, the phase $\nu \in \mz_8$ has $\nu$ right-moving Majoranas charged under $\mz_2$ and $\nu$ left-moving Majoranas neutral under $\mz_2$. For even $\nu$, we can group $\nu$ chiral Majoranas into $\nu/2$ chiral complex (Dirac) fermions ($c  =1$). 

We now proceed to 3+1D. As already noted, in 3+1D modulo bosonic SPTs there is a single non-trivial fermionic SPT with internal symmetry ${\mathbb Z}_2 \times {\mathbb Z}_4 \times {\mathbb Z}^f_2$.\cite{Kapustin2017,CTW}  This phase is the generator ($n =1$) of a $\Z_4$ subgroup in the classification, where the $n = 2$ phase is a bosonic SPT. There is also one other root bosonic SPT with this symmetry,\cite{WL2015} bringing the full classification to $\Z_4 \times \Z_2$.\cite{JuvenSpin} Here we focus on the intrinsically fermionic root phase. 
A physical picture of this phase consists of a soup of $\Z_4$ domain walls. Each such domain wall is decorated \cite{Chen2013b} with the $\nu  = 2$ $\mz_2 \times \mz^f_2$ 2+1D fermion SPT. 

A key property of SPTs in both 2+1D and 3+1D is the braiding statistics of flux defects \cite{Levin_Gu, Wang2014a, WL2015}. For $\nu  = 2$ $\mz_2 \times \mz^f_2$ SPT in 2+1D, we have\cite{Gu_Levin}
\beq 2 \theta_{\mfg_1} = \frac{\pi}{2}, \quad 2 \theta_{\mfg_1,\mfg_1} = \pi, \quad 2 \theta_{\mfg_1, \mfg_f} = \pi \label{Z2braid} \eeq  
Here, $\theta_i$ denotes the self-statistics of a flux defect $i$, and $\theta_{i,j}$ denotes the full-braid mutual statistics of  flux defects $i$ and $j$. $\mfg_1$ labels the generator of the $\mz_2$ symmetry and $\mfg_f$ the generator of $\mz^f_2$ symmetry. (All quantities are given modulo $2\pi$ and the extra factors of $2$ on the left-hand-side are necessary to eliminate the dependence coming from fusing a defect with point charges).

In 3+1D, the relevant process is the three loop braiding \cite{Wang2014a}, i.e. braiding of two loops $i$, $j$ linked with a third loop $k$. For the  intrinsically fermionic ${\mathbb Z}_2 \times {\mathbb Z}_4 \times {\mathbb Z}^f_2$ SPT, we have\cite{CTW}
\bea &&2 \theta_{\mfg_1; \mfg_2} =  \frac{\pi}{2}, \quad 2 \theta_{\mfg_1, \mfg_1; \mfg_2} = \pi, \quad 2 \theta_{\mfg_1, \mfg_f; \mfg_2} = \pi, \nonumber \\
 &&4 \theta_{\mfg_1, \mfg_2; \mfg_1} = \pi. \label{eq:3loop}\eea
with all other braiding phases being trivial. Here $\theta_{i; j}$ is the self-statistics of loop $i$ linked with a base loop  $j$, and $\theta_{i, j; k}$ is the full-braid mutual statistics of loops $i$, $j$ linked with a base loop $k$. Furthermore, $\mfg_1$ is the generator of $\mz_2$, $\mfg_2$ is the generator of $\mz_4$, and $\mfg_f$ is again the generator of $\mz^f_2$. Focusing on the braiding of $\mz_2$ and $\mz^f_2$ fluxes in the presence of a $\mz_4$ base flux, we see that it exactly matches the braiding in the $\nu = 2$ $\mz_2 \times \mz^f_2$ 2+1D SPT, Eq.~(\ref{Z2braid}). So we can think of the surface of $\mz_4$ flux loops as decorated with  $\nu  =2$ $\mz_2 \times \mz^f_2$ SPT; in particular, the loop traps gapless modes, which coincide with those of the $\nu =2$ SPT edge.  Here is another consequence of the above braiding statistics: let us compactify the system on a spatial manifold $M_2 \times S^1$, where $M_2$ is a two-dimensional manifold. We can view the whole system as a 2+1D SPT with $\mz_2 \times \mz^f_2$ symmetry. Then switching the $\mz_4$ flux along $S^1$ from $0$ to $\mfg_2$ changes the 2+1D SPT index by $\Delta \nu  =2$. 

\begin{figure}
\begin{center}
\includegraphics[width =3.2in]{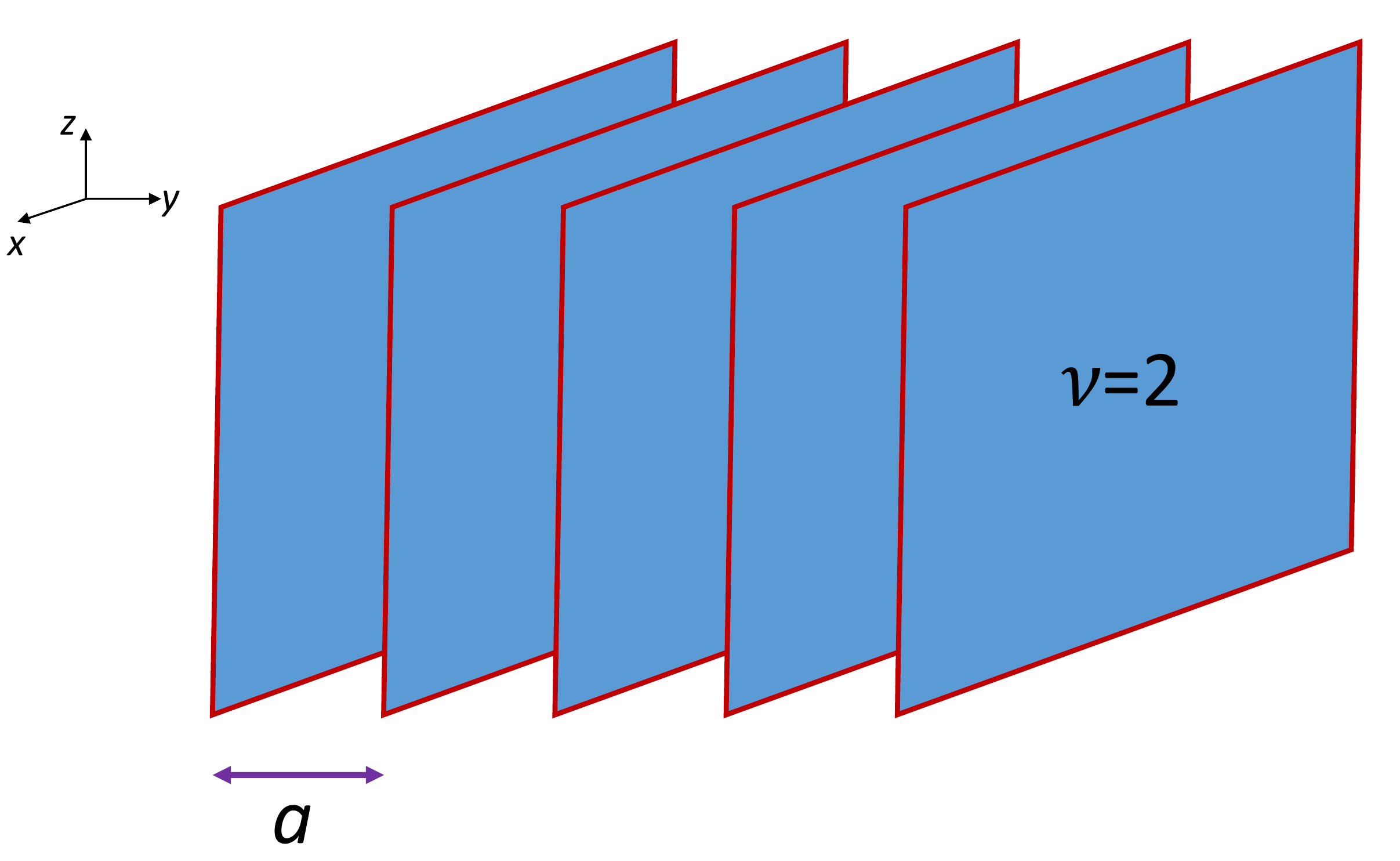} 
\end{center}
\caption{A 3+1D crystalline SPT with $\mz_2 \times \mz \times \mz^f_2$ symmetry constructed by stacking layers of 2+1D $\nu  =2$ $\Z_2 \times \Z^f_2$ SPTs along the $y$ direction. Brown lines denote  gapless edge modes.}
\label{fig:stack}
\end{figure}

\begin{figure}
\begin{center}
\includegraphics[width =1.8in]{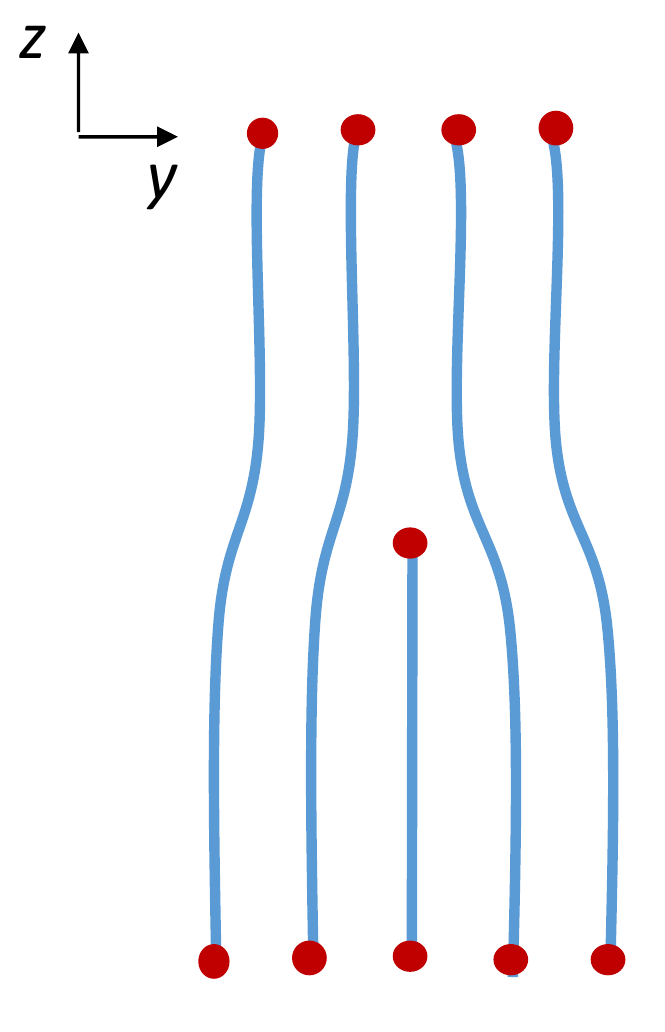} 
\end{center}
\caption{A dislocation in the crystalline SPT of fig. \ref{fig:stack} hosts a $\nu=2$ $\mz_2 \times \mz^f_2$ SPT edge mode. The brown circles denote these edge modes propagating perpendicular to the plane $yz$ plane.}
\label{fig:dislocation}
\end{figure}

We now proceed to describe the closely related 3+1D crystalline SPT protected by ${\mathbb Z}_2 \times {\mathbb Z}\times {\mathbb Z}^f_2$ symmetry. We choose $\mz$ to be translation by lattice constant $a$ along the $y$ direction. We build the phase by stacking along the $y$ direction layers of $\nu = 2$ $\mz_2 \times \mz^f_2$ 2+1D SPT lying in the $xz$ plane, see figure \ref{fig:stack}. This clearly defines a non-trivial crystalline SPT. Indeed, let us make the $y$ direction periodic. For the length of the $y$ direction large, but finite, we can think of the system as being effectively two-dimensional lying in the $xz$ plane. Then changing the number of layers (i.e. the length of the $y$ direction) by one changes the $\mz_2 \times \mz^f_2$ SPT index of this two-dimensional system by $\Delta \nu  =2$. This means that the crystalline SPT is non-trivial. For instance, the surfaces possessing translational symmetry along the $y$ direction (e.g. the $z  =0$ surface) cannot be trivially gapped, since for $L_y \neq 0 \,\,({\mathrm{mod}}\,4\,)\,\,$ they are effectively the edge of a non-trivial 2+1D SPT. Now, there is an evident similarity to the ${\mathbb Z}_2 \times {\mathbb Z}_4 \times {\mathbb Z}^f_2$ SPT described above, where  inserting a $\mz_4$ flux along a cycle resulted in $\Delta \nu = 2$. For the present crystalline SPT the same occurs when we change $L_y \to L_y + 1$. But for translational symmetry, changing the length of the system by $1$ is, indeed, the definition of inserting a $\mz$ flux. So the crystalline SPT behaves like the internal symmetry SPT above with $\mz_4 \to \mz$. A slightly more physical way to think about the same issue is to consider ``flux defects" of the translational symmetry $\mz$, i.e. the dislocations. These correspond to one of the layers terminating prematurely in the bulk - thus, the dislocation obviously carries the edge-modes of a 2+1D $\nu = 2$ ${\mathbb Z}_2 \times  {\mathbb Z}^f_2$ SPT - see figure \ref{fig:dislocation}.

\section{Surface topological order for ${\mathbb Z}_2 \times {\mathbb Z}_4 \times {\mathbb Z}^f_2$ SPT.}
\label{sec:STO}
Here we briefly describe our conjecture for the STO of the 3+1D ${\mathbb Z}_2 \times {\mathbb Z}_4 \times {\mathbb Z}^f_2$ SPT. For the crystalline ${\mathbb Z}_2 \times {\mathbb Z} \times {\mathbb Z}^f_2$ SPT, we will explicitly construct such a STO termination in section \ref{sec:stack}. In the present section, we will give a physical argument why such a topological order is anomalous both for the case of internal and  crystalline symmetries.

Our proposed STO has anyon content $\mz_4   \times \{1, f\}$. The first factor is a $\mz_4$ gauge theory ($\mz_4$ version of the toric code). The second factor  contains only one non-trivial particle - the physical fermion $f$. We remind the reader that the $\mz_4$ gauge theory  is generated by Abelian anyons $e$ and $m$, where $e$ and $m$ are both self-bosons, and the full-braid mutual statistics $\theta_{e,m} = e^{ 2\pi i/4}$. Further, $e^4 = m^4 = 1$. 

The anomaly of the STO comes from the combination of anyon permutation under the $\mz_2 \times \mz_4$ symmetry and the fractional quantum number assignments. We take the $\mz_2$ symmetry not to permute the anyons, while the generator of $\mz_4$ acts as:
\beq \mz_4:\,\, e\to e, \quad m \to m e^2 f \label{eq:Z4perm}\eeq
It is easy to check that this permutation preserves the fusion and braiding rules. Next, we specify the symmetry fractionalization on the anyons. We take $e$ to carry charge $1/4$ under $\mz_2$, while  $m$  carries no fractional charge under $\mz_2$. Schematically, we write,
\beq \mz_2: e \to e^{\pi i/4} e, \quad m \to m \label{eq:Z2frac}\eeq
Note that we have not specified the fractionalization data for the $\mz_4$ symmetry here. This data will be discussed in section \ref{sec:H3comp} and explicitly derived for the crystalline SPT (with $\mz_4 \to \mz$) in section \ref{sec:stack}. For the crystalline SPT our proposed STO has the same anyon content and symmetry data (\ref{eq:Z4perm}), (\ref{eq:Z2frac}), with the replacement $\mz_4 \to \mz$.

We now argue that the topological order with the symmetry action above is anomalous, i.e. it cannot exist strictly in 2+1D.  To see this, imagine introducing flux defects $a$ of the $\mz_2$ symmetry. By potentially pasting on $p+ip$ superconductors and $\mz_2 \times \mz^f_2$ SPTs, we can ensure that the flux defects are Abelian\footnote{Indeed, consider the topological order with the fermion parity gauged. There are 16 such modular extensions, corresponding to the 16 modular extensions of $\{1, f\}$. By pasting on a suitable number of $\mz_2 \times \mz_4$ symmetric $p+ip$ superconductors (e.g. constructed from fermions neutral under $\mz_2 \times \mz_4$) we can ensure that the modular extension is the toric code $\{1,x, y, f\}$, with $x$, $y$ - the fermion parity fluxes. Since the original anyons are not permuted by $\mz_2$, consistency with fusion/braiding requires that under $\mz_2$ either $x \to x, \, y \to y$ or $x \leftrightarrow y$. In the second case, we can paste on an $m  =1$ $\mz_2 \times \mz^f_2$ SPT. When the fermion parity in such an SPT is gauged, one gets the toric code $\{1, x, y, f\}$ with $\mz_2: x \leftrightarrow y$. So, pasting  on this SPT makes $x \to x$ and $y \to y$ under $\mz_2$. Then none of the anyons in the modular extension are permuted by $\mz_2$, which means that the $\mz_2$ defect must be Abelian.}. We can think of the $\mz_2$ symmetry action on $e$ as arising from braiding $a$ around $e$, i.e. such a process results in a Berry phase $\frac{\pi}{4}$. Similarly, braiding $a$ around $m$ should give no Berry phase. Now, since $a$ is a $\mz_2$ flux, fusing $a \times a$ should give one of the original anyons in the theory. But braiding $a \times a$ with $e$ should give Berry phase $\frac{2 \pi}{4}$ and braiding $a \times a$ with $m$ should have given Berry phase $0$. The only anyons with the right braiding properties are $m$ and $mf$, i.e. $a \times a = m$ or $a \times a = mf$. Let us focus on the case $a \times a = m$, the argument for $a \times a = mf$ is identical. Now, under the $\mz_4$ action the $\mz_2$ flux defect $a$ can transform to a different $\mz_2$ flux defect $b$. All the $\mz_2$ flux defects have to be related by fusion with an anyon in the original theory, so $b  = a \times c$ for some  $c \in \mz_4 \times \{1, f\}$. Now, for consistency of $\mz_4$ symmetry with fusion $b \times b = m e^2 f$, which means that $c \times c = e^2 f$. But there is no anyon in $\mz_4 \times \{1, f\}$ which fuses with itself  to $e^2 f$. So we've arrived at a contradiction and the topological order is anomalous.

We note that the above argument for the anomaly would likewise go through if the $\mz_4$ internal symmetry is replaced by the $\mz$ translation symmetry. Indeed, we have not attempted to introduce defects of $\mz_4$ symmetry anywhere in the argument and only used its global action.

\section{Surface topological order for the $\Z_2 \times \Z \times \Z^f_2$ crystalline fermion SPT.} \label{sec:stack}

\begin{figure}
\begin{center}
\includegraphics[width =1.8in]{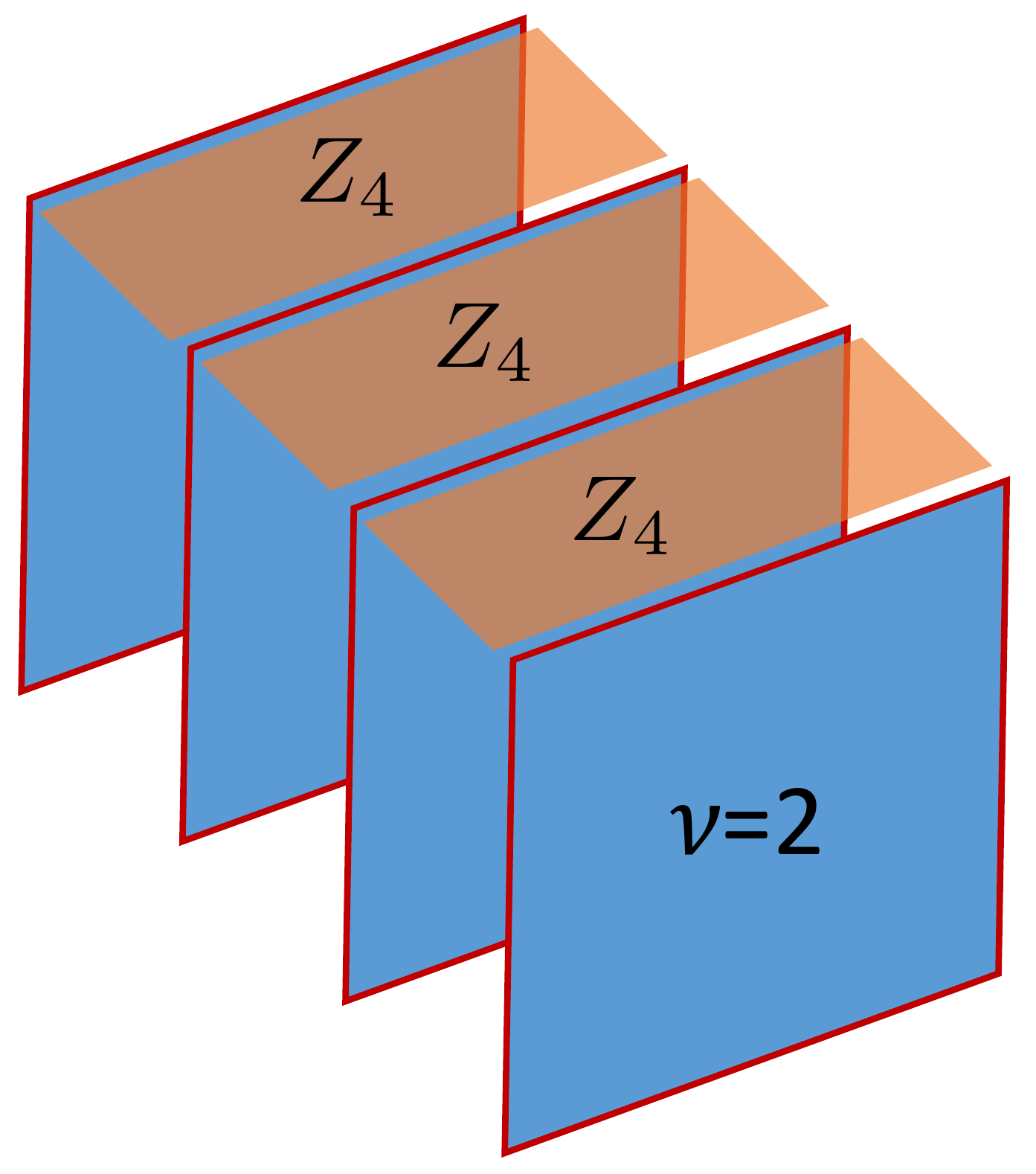}\quad \includegraphics[width =1.8in]{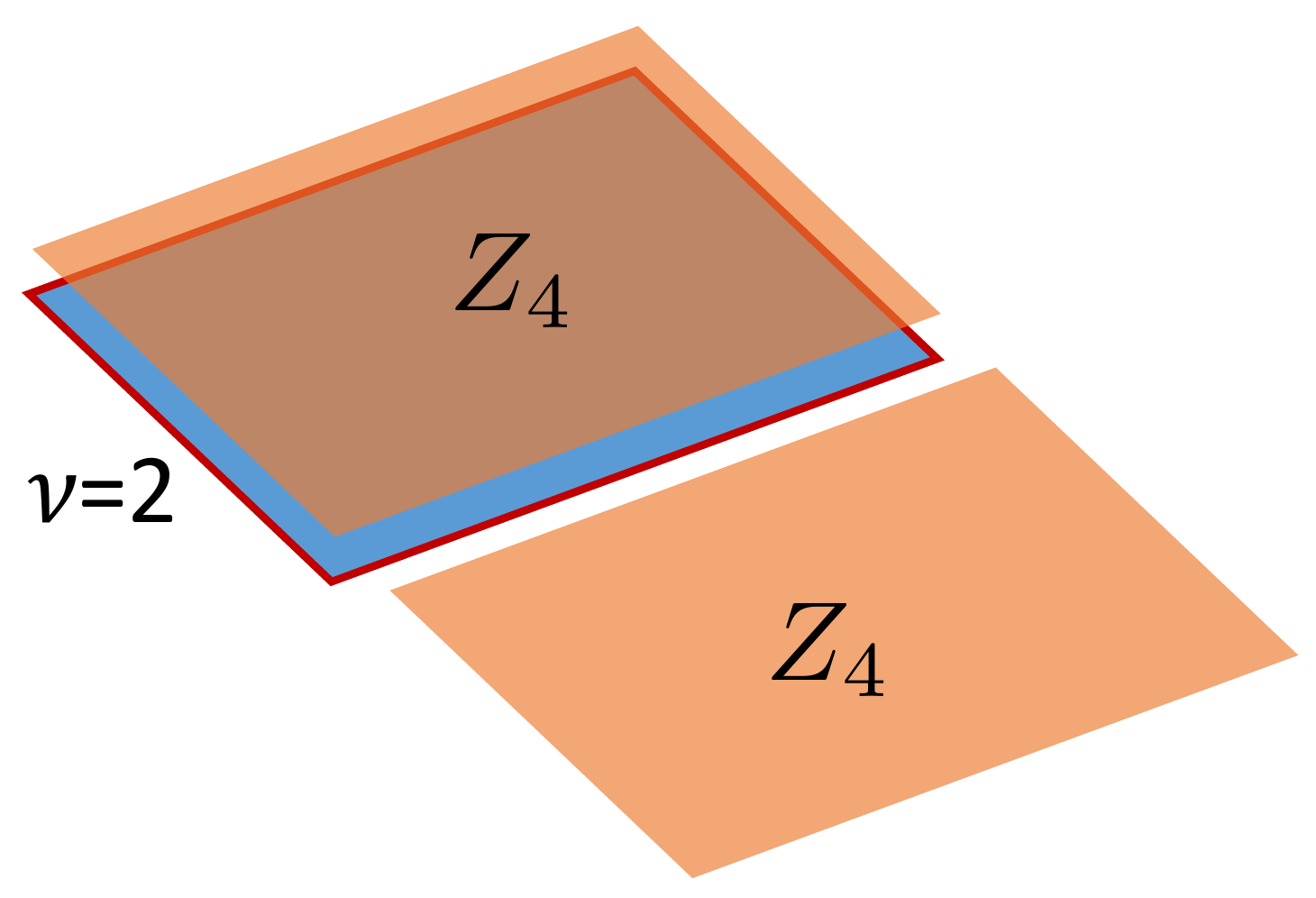}
\end{center}
\caption{{Top: Constructing a topological order for the $z = 0$ surface of the crystaline SPT in Fig. \ref{fig:stack}. The surface area between  neighboring $\mz_2 \times \mz^f_2$ SPT edges is decorated with a strip of $\Z_4$ topological order (orange). These strips are then ``stitched" together to produce a macroscopic $\Z_4$ topological order and gap out the modes associated with the $\mz_2 \times \mz^f_2$ SPT edges. Bottom: A thought experiment illuminating why the stitching procedure works: a $\Z_4$ gauge theory stacked with a $\nu = 2$ $\Z_2 \times \Z^f_2$  SPT (left) is, in fact, in the same $\mz_2 \times \mz^f_2$ SET phase as the $\Z_4$ gauge theory without the additional SPT (right).}}
\label{fig:stackdec}
\end{figure}

In this section we explicitly construct the STO discussed above on the surface of the 3+1D $\Z_2 \times \Z \times \Z^f_2$ crystalline fermion SPT. 
We start with the ``stack" model of the crystalline SPT (figure \ref{fig:stack}) discussed in section \ref{sec:bulk}. Consider e.g. the $z = 0$ surface of the stack: it consists of an array of 1+1D gapless $\nu = 2$ SPT edges. Our goal is to gap out these edges without breaking the symmetry. Let us decorate the surface with strips of $\mz_4$ topological order arranged periodically, see figure \ref{fig:stackdec} (top).
We choose the $\mz_2$ symmetry to act on this topological order as in Eq.~(\ref{eq:Z2frac}). We would like to ``stitch" the strips of $\mz_4$ topological order together and in the process gap out the  SPT edges. {To understand why this stitching works, we can focus on a single T-junction on the surface and perform the following thought experiment:} imagine ``bending" the SPT layer onto the surface, figure \ref{fig:stackdec} (bottom). On the right of the junction we then have the $\mz_4$ topological order and on the left - the same topological order stacked with a $\nu =2$ SPT.  Now, we make a key claim: the $\mz_4$ topological order can ``absorb" the $\nu =2$  SPT, i.e. the  state on the right of the junction and the state on the left of the junction are identical as $\mz_2 \times \mz^f_2$ SETs. Therefore, the two can be stitched together without breaking the $\mz_2$ symmetry. Performing this stitching on the entire array of T-junctions we obtain the desired STO.

We now elaborate on the above steps.

 \vspace{0.45cm}


{\it Absorbing the $\nu = 2$ $\Z_2 \times \Z^f_2$ SPT}

Let us show that the $\Z_4$ gauge theory with $\Z_2$ action (\ref{eq:Z2frac}) can absorb a $\nu  =2$ $\Z_2\times \Z^f_2$ SPT. Here, we imagine a purely 2+1D setting with a 1+1D edge (left side of figure \ref{fig:stackdec}, bottom).  Recall, the edge of the $\nu = 2$ SPT admits a Luttinger liquid description:
\begin{align} L^{edge}_{\nu =2} = \frac{1}{4\pi} (\d_x \nu_R \d_t \nu_R - \d_x \nu_L \d_t \nu_L) \end{align}
where  $e^{i \nu_{R/L}}$ is the electron. We drop kinetic energy terms of form $(\d_x \nu_{R/L})^2$ here and below.
Likewise, we can describe the edge of the $\mz_4$ gauge theory by:
\beq L^{edge}_{\Z_4} = \frac{4}{2\pi} \d_{x} \phi \d_t \theta \eeq
where $e^{i \phi}$ is the $e$ particle, and $e^{i \theta}$  - the $m$ particle. Under $\Z_2$,
\beq \Z_2: \,\, \phi \to \phi + \frac{\pi}{4},  \quad \nu_R \to \nu_R + \pi \label{Z2edge}\eeq
with $\theta$ and $\nu_L$ unaffected. Combining the STO edge and the SPT edge,
\beq L^{edge} = \frac{1}{4\pi} \d_{x} \Phi_I K_{IJ} \d_{t} \Phi_J \eeq
where $\Phi = (\theta, \phi, \nu_R, \nu_L)$, 
\beq K = \left(\begin{array}{cccc} 0 & 4 & 0 &0\\ 4&0&0&0\\0 &0&1&0\\0 &0&0&-1\end{array}\right)\eeq
and
\beq \Z_2: \,\, \Phi \to \Phi + \pi (0,\frac14,1, 0)\eeq
Now, consider a $SL(4,\Z)$ change of variables
\beq \Phi' = S \Phi, \quad S =  \left(\begin{array}{cccc} 1 & -2 & 1 &0\\ 0&1&0&0\\0 &4&-1&0\\0 &0&0&1\end{array}\right) \label{eq:Sdef}\eeq
We have $S^2 = 1$ and $S^T K S = K$. This means that $S:\,\, m \to me^2f, \,\, e \to e$, is a $\Z^S_2$ symmetry of the $\Z_4 \times \{1,f\}$ topological order (we use the superscript $S$ to differentiate it from the original $\Z_2$ symmetry). However, in our edge construction, $S$ does not commute with the $\Z_2$ symmetry (\ref{Z2edge})\footnote{This is the first hint that the transformation $S: m \to m e^2 f, e \to e$ has a mixed anomaly with the $\Z_2$ symmetry.}.
In fact, 
\beq \Z_2: \Phi' \to \Phi' + \pi (\frac12, \frac14,0,0) \eeq
The transformation $\Z_2: \theta' \to \theta' + \frac{\pi}{2}$ is a pure gauge rotation, so can be ignored. Thus,
\beq \Z_2: \Phi' \to \Phi' + \pi (0,\frac14, 0,0)\eeq
This is the same as in our original $\Z_4$ topological order, but without the extra $\nu = 2$ SPT on top. Indeed, the extra $\nu'_R$, $\nu'_L$ modes can be gapped out by
\beq \delta L = - \lambda \cos(\nu'_R - \nu'_L) = - \lambda \cos (4 \phi - \nu_R -\nu_L) \label{gapnup}\eeq

When written in the primed variables it is especially clear that this gapping process does not break the $\mz_2$ symmetry.
This proves that the $\Z_4$ gauge theory can absorb the $\nu = 2$ SPT.


\vspace{0.25cm}

{\it Stitching.}

\begin{figure}
\begin{center}
\includegraphics[width = 3.5in]{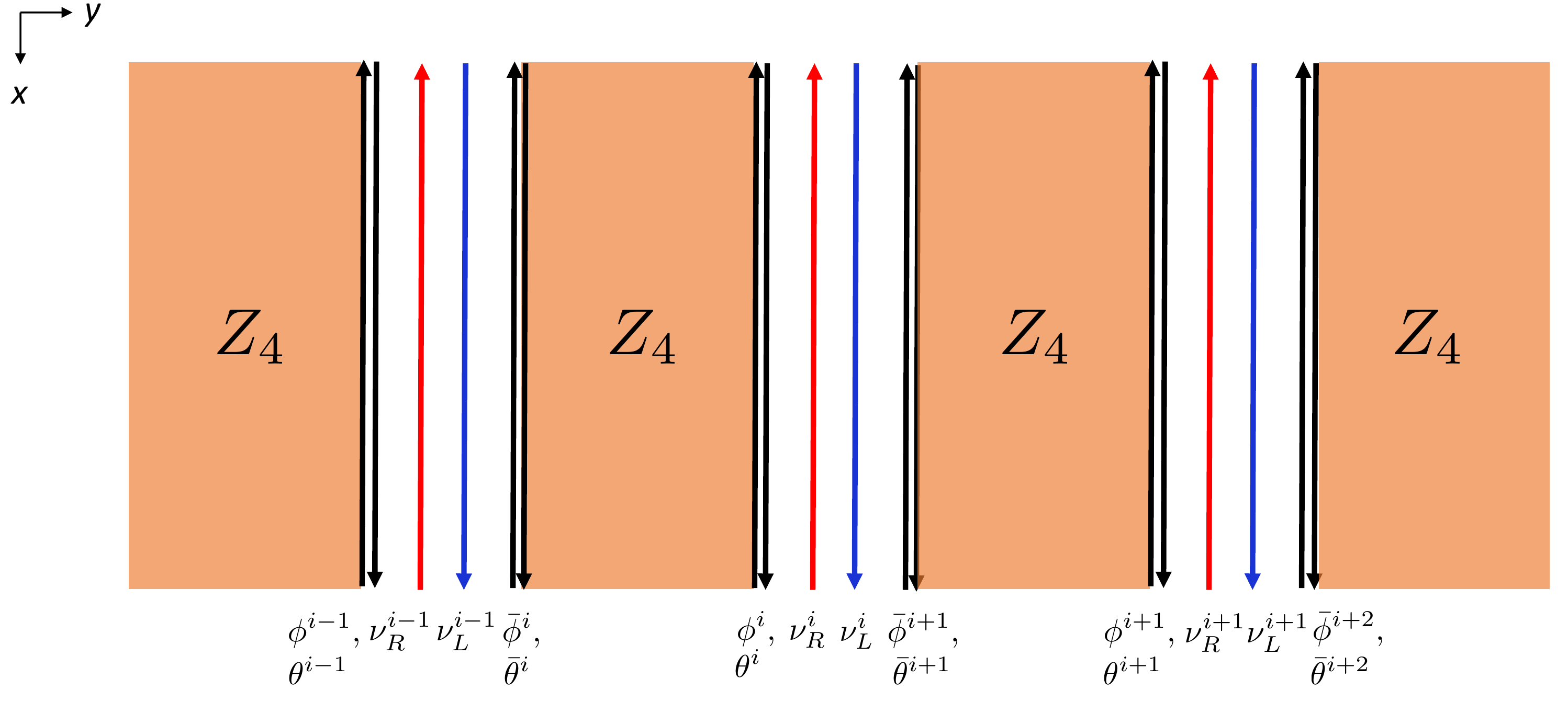}
\end{center}
\caption{
{Expanded top view of the surface decoration in figure \ref{fig:stackdec} (top). $\nu^i_R$ (red, $\Z_2$ charged) and $\nu^i_L$ (blue, $\Z_2$ neutral) are edge modes of $\nu = 2$ $\Z_2 \times \Z^f_2$ SPTs in the stack. 
$\phi^i$, $\theta^i$, $\bar{\phi}^i$, $\bar{\theta}^i$ are the edge states associated with strips of $\Z_4$ topological order. The terms in Eq. (\ref{gap3}) gap out all edge modes without breaking the symmetry and sew together the $\Z_4$ topological orders. }}
\label{fig:surfdec}
\end{figure}


We are now ready to stitch the strips of topological order in figure \ref{fig:stackdec} (top) together and eliminate the gapless SPT modes. A top view of our construction is shown in figure \ref{fig:surfdec}. The array of gapless modes coming from the $\nu = 2$ SPTs is described by:
\beq L^{surface}_{\nu = 2} = \frac{1}{4\pi} \sum_i (\d_x \nu^i_R \d_t \nu^i_R - \d_x \nu^i_L \d_t \nu^i_L)\eeq
where $i$ labels the $y$ coordinate of the mode, $y = i a$. Under translations, 
\beq \Z: \nu^i_{R/L} \to \nu^{i+1}_{R/L}\eeq
and under $\Z_2$, 
\beq \Z_2: \nu^i_{R} \to \nu^i_R +\pi, \,\, \nu^i_L \to \nu^i_L\eeq
The strips of $\Z_4$ topological order are described by:
\beq L^{surface}_{\Z_4}  = \frac{4}{2\pi} \sum_i (\d_{x} \phi^i \d_t \theta^i - \d_x \bar{\phi}^i \d_t \bar{\theta}^i) \eeq
The unbarred and barred variables correspond to right and left edges of the topological order strips, i.e. $\phi^i, \theta^i$ live at $y = i a$, and $\bar{\phi}^i, \bar{\theta}^i$ at $y = (i-1) a$. As before, the $\Z_2$ charge assignments are
\beq \Z_2: \phi^i \to \phi^i + \frac{\pi}{4},\,\, \bar{\phi}^i \to \bar{\phi}^i + \frac{\pi}{4} \eeq
and translations $\Z$ act by shifting the index $i \to i + 1$. 

If we focus on all the modes at $y = i a$, we have
\begin{align}
 L_i &= \frac{1}{4\pi} (\d_x \nu^i_R \d_t \nu^i_R - \d_x \nu^i_L \d_t \nu^i_L) \nonumber \\ &+ \frac{4}{2\pi}(\d_{x} \phi^i \d_t \theta^i - \d_x \bar{\phi}^{i+1} \d_t \bar{\theta}^{i+1})
 \end{align}
Following the discussion in the previous section, we consider the gapping term,
\begin{align}
 \delta L_i &=  -\lambda_1 \cos (4 \theta'^i - 4 \bar{\theta}^{i+1}) - \lambda_2 \cos(4 \phi'^i -  4 \bar{\phi}^{i+1})\nonumber \\ & -\lambda_3 \cos(\nu'^i_R - \nu'^i_L) \label{gap3pr} \\=
&-\lambda_1 \cos(4 (\theta^i - \bar{\theta}^{i+1}) - 8 \phi^i +4 \nu^i_R) \nonumber \\
&- \lambda_2 \cos(4 (\phi^i - \bar{\phi}^{i+1})) - \lambda_3 \cos(4 \phi^i - \nu^i_R - \nu^i_L) \label{gap3}
\end{align}
where as before $(\theta^i, \phi^i, \nu^i_R, \nu^i_L)' = S (\theta^i, \phi^i, \nu^i_R, \nu^i_L)$, Eq.~(\ref{eq:Sdef}). The terms above gap out all modes at $y = ia$, as is especially clear in the primed variables, Eq.~(\ref{gap3pr}). Furthermore, they effectively glue $e^{i \bar{\phi}^{i+1}} \sim e^{i \phi'^{i}} = e^{i \phi^i}$, $e^{i \bar{\theta}^{i+1}} \sim e^{i \theta'^{i}} = e^{i \theta^i} e^{-2 i \phi^i} e^{i \nu^i_R}$. This means that both the $e^{i \phi}$ and $e^{i \theta}$ particles can move across the ``seams" at $y = i a$, i.e. the surface as a whole forms a single $\Z_4 \times \{1, f\}$ topological order. 
To label the topological charges globally, we will use, say, the $i = 0$ strip as the reference point, so  $e^{i \theta^0}$ is $m$. Since $e^{i \phi^i} \sim e^{i \phi^{i+1}}$ in the topological sense, any $e^{i \phi^i}$ is in the $e$ sector.

\vspace{0.25cm}
{\it $\Z_2 \times \Z$ action.}

Now, let us investigate the action of $\Z_2 \times \Z$ symmetry on the above STO. The $e$ particle carries $\Z_2$ charge $\frac14$ throughout. The $m$ particle carries no fractional $\Z_2$ charge. Under translation,
\begin{align}
\Z:\,\quad& e = e^{i \phi^i} \to e^{i \phi^{i+1}} \sim e \\
&m = e^{i \theta^0} \to e^{i \theta^1} \sim e^{i \bar{\theta}^1} \sim e^{i \theta^0} e^{-2 i \phi^0} e^{i \nu^0_R} \sim m e^2 f
\end{align}
This is exactly the action (\ref{eq:Z4perm}).

For topological orders with a combination of an internal symmetry $G_{int}$ ($G_{int} = \Z_2$ in our case) and translation $\Z$ there is one other piece of data that characterizes the symmetry fractionalization: whether the action of $G_{int}$ and $\Z$ commutes on the anyons. Generally, on an anyon $T_s g_{int} T^{-1}_s g^{-1}_{int}= e^{i \varphi}$ for some phase $e^{i \varphi}$, where $g_{int} \in G_{int}$ and $T_s$ is a translation by $s$ lattice sites. To extract the phases $e^{i \varphi}$ we follow the procedure of Ref. \onlinecite{Cheng_Bonderson}.  Imagine a state $|b, y\rangle$ with an anyon $b$ at position $y$ (there is another anyon $\bar{b}$ at $y = -\infty$ that goes along for the ride). Under translation by $s$,  $T_{s} |b, y\rangle = W_b(y, y+s) |b, y\rangle$ where $W_b(y,y+s)$ is a string operator that moves $b$ from $y$ to $y+s$. Crucially, we assume here that $T_s$ preserves the anyon type, $T_s:\,\, b \to b$ (otherwise, the string operator does not exist). By computing the charge of $W_b(y,y+s)$ under $\Z_2$, we find if $\Z_2$ and $T_s$ commute or anti-commute on $b$. Let's implement this procedure. We begin with the anyon $e$. We have,
\begin{align}
 T_{a} e^{i \phi^i} |0\rangle &= e^{i \phi^{i+1}} |0\rangle \sim e^{i \phi^{i+1}} e^{-i \bar{\phi}^{i+1}} e^{i \phi^i} |0\rangle \nonumber \\ &= W_e(i,i+1) e^{i \phi^i}|0\rangle 
\end{align}
with $W_e(i,i+1) = e^{i (\phi^{i+1} - \bar{\phi}^{i+1})}$. Here we used the fact that due to the gapping terms, $e^{-i \bar{\phi}^{i+1}} e^{i \phi^i} |0\rangle \sim |0\rangle$. $W_e$ is neutral under $\Z_2$, so $T_a$ and $\Z_2$ commute on $e$. Now, $m$ does not preserve anyon type under translation by $a$, however, $m^2$ does. We have,
\begin{align} T_{a} e^{2 i \theta^0} |0\rangle &= e^{2 i \theta^1} |0\rangle \nonumber \\ &\sim e^{2 i \theta^1} e^{-2 i \bar{\theta}^1} e^{-4 i \phi^0} e^{2 i \nu^0_R} e^{2 i\theta^0} |0\rangle \nonumber \\ &= W_{m^2}(0,1) e^{2 i \theta^0} |0\rangle  \end{align}
with $W_{m^2}(0,1) = e^{2 i \theta^1} e^{-2i \bar{\theta}^1} e^{-4 i \phi^0} e^{2 i \nu^0_R}$. $W_{m^2}(0,1)$ is odd under $\Z_2$, so $\Z_2$ and $T_a$ anti-commute on $m^2$. Finally, we observe that while $m$ is not invariant under $T_a$, it is invariant under $T_{2a}$. We have
\begin{align} T_{2 a} e^{i \theta^0}|0\rangle &= e^{i \theta^2}|0 \rangle \sim e^{ i \theta^2} e^{-i \bar{\theta}^2} e^{i \theta^1} e^{-2 i \phi^1} e^{i \nu^1_R}|0\rangle \nonumber \\ &= W_{m}(0,2) e^{i \theta_0} |0\rangle \end{align}
with 
\begin{align}
W_m(0,2) = e^{i (\theta^2- \bar{\theta}^2)} e^{i (\theta^1 - \bar{\theta}^1)} e^{-2i (\phi^1 -\bar{\phi}^1)} e^{-4 i \phi^0} e^{i \nu^1_R} e^{i \nu^0_R}.
\end{align}
$W_{m}(0,2)$ is odd under $\Z_2$, so $\Z_2$ and $T_{2a}$ anticommute on $m$. 

We summarize the above data in table \ref{proj}.

\begin{table}[t]
\beq
\begin{array}{|c|c|c|c|}
\hline
   & \mfg^4_1& [\mfg_1,T_a]  & [\mfg_1, T^2_a]  \\ \hline
e   & -1& 1&  1       \\ \hline
m   &1&     & -1  \\ \hline
m^2  &1&   -1 & 1    \\ \hline 
\end{array}
\nonumber
\eeq
\caption{Symmetry fractionalization in the STO of $\Z_2\times \Z \times \Z^f_2$ crystalline topological insulator. Here, $\mfg_1$ is the generator of $\Z_2$ and $T_a$ is the generator of translations $\Z$, and $[g, h] = g h g^{-1} h^{-1}$. The fractionalization is not well-defined for the blank entry.}
\label{proj}
\end{table}

Let us describe an intuitive understanding of the origin of the anyon permuting symmetry.  Our aim is to stitch together the adjacent $\Z_4$ topological orders at the bottom of Figure \ref{fig:stackdec}. Maintaining $\Z_2$ symmetry implies that the corresponding $\Z_2$ symmetry flux can freely tunnel across the interface. Let the $\Z_2$ symmetry flux on the right half be called $v_R$: this is a boson which satisfies $v_R^2=m_R$ so that the $e$-particles of the $\Z_4$ topological order carry  $\Z_2$ charge $1/4$. Next, let us call the symmetry flux on the left side of the junction $v_L$. The key observation is that on passing through the fermion SPT layer, the $\Z_2$ symmetry flux acquires a topological spin of $e^{i\pi/4} $. Thus $v_L^2 $ is a fermion, and the only consistent assignment is $v_L^2=m_L f$.  For the $\Z_2$ symmetry fluxes to smoothly tunnel across the junction, we need to condense $v_L^{-1}v_R$ on the boundary, which is disallowed due to the mismatch in topological spin. To rectify this, there is a different combination using the modified flux: $v'_L = e^{-1}_Lv_L$, which allows for a condensate of $({v}_L')^{-1}v_R$ on the boundary. This implies the square is also condensed, which from the previous relations tells us that $m_R \sim e_L^2m_Lf$. This clarifies why a $\Z_4$ topological order is required (and a simpler $\Z_2$ toric code order would not suffice).  Also note, this phenomenon of symmetry permuting anyons is {not necessarily required for STOs of boson SPTs. For instance, take the ``double," $n =2$, of the presently considered $n =1$ fermionic SPT. We would use the $\nu=4$ $\Z_2 \times \Z^f_2$ SPT in the stack construction, which is essentially bosonic. Now the topological spin of the flux defect changes only by $e^{i\pi/2}$, which is not visible on considering $v^2$.} 

Before we conclude this section we note that there is a simple generalization of the bulk SPT and surface STO discussed here to the case when the $\Z^f_2$ symmetry is enlarged to a full $U(1)$ particle-number symmetry. We discuss this generalization in appendix \ref{app:U1}. A nice feature of the case with the $U(1)$ symmetry is that the projective transformation of $m^2$ under $\Z_2 \times \Z$ on the surface is closely linked to the identical projective transformation of the $U(1)$ monopole in the crystalline SPT bulk. In addition, this gives us a promising direction to the physical realization of this crystalline SPT state. The symmetries required are charge conservation, i.e. $U(1)$, a translation symmetry $\Z$ and an additional internal $\Z_2$ symmetry. Furthermore, time reversal must be broken as in a magnetic insulator. Finally we note that the $\Z_2$ symmetry cannot be lifted to a second $\Z$  translation symmetry, since the symmetry fractionalization relies on the finiteness of this part of the symmetry group.

\section{General anomaly matching condition for super-cohomology fermionic SPTs with onsite symmetries.}
\label{sec:H3f}

We now describe a general anomaly matching condition between bulk fermionic SPT orders and topologically ordered surface states, at an algebraic level.

\subsection{Bulk fermionic SPT order}

First, let us recall the classification of 3+1D fermionic SPTs with symmetry group $G\times \Z^f_2$ proposed in Refs. [\onlinecite{Kapustin2017, Wang_Gu}].  Modulo stacking bosonic SPTs, such fermionic SPTs are classified by $\Z_2$-valued functions $\rho(f,g,h)$ and $\sigma(f,g)$ satisfying the properties:

\begin{align}
d \rho = \sigma \cup \sigma \\
d \sigma = 0
\end{align}
This data is also subject to the following redundancy conditions:
\begin{align}
\sigma &\rightarrow \sigma + d \lambda \\
\rho &\rightarrow \rho + d \beta + \lambda \cup d \lambda + d \lambda \cup_1 \sigma
\end{align}
For a definition of the cup product, higher cup product, and the derivative $d$, see e.g. Ref. [\onlinecite{Kapustin2017}].  There is a further condition on $\sigma$, $\rho$, which in the simplest case $\sigma = 0$ reads $[\frac{1}{2} \rho \cup_1 \rho] = 0$ as a cohomology class in $H^5(G,\mathbb{R}/\Z)$.\cite{Wang_Gu, Kapustin2017}

Phases with non-trivial $\sigma \in H^2(G, \Z_2)$ can be visualized in the following way. As with all symmetric phases, we can think of the ground state as a soup of domain-walls of $G$. Consider a 1d junction of three domain walls corresponding to elements $g$, $h$ and $gh$. If $\sigma(g,h) \neq 0$, such a junction traps a Kitaev chain. Phases with non-zero $[\sigma]$ are called `beyond super-cohomology' phases, and are not going to be the focus of this paper, although we will briefly speculate about the surface topological orders that they can support in the next subsection.

The super-cohomology phases, i.e. the ones with $[\sigma]=0$, will be the focus of this paper.  In this case, $\sigma$ itself can be taken to be $0$ by performing the appropriate gauge transformation.  The above data then just reduces to a cohomology class $[\rho] \in H^3(G,\Z_2)$ - this is the bulk SPT invariant that we will try to match onto a surface anomaly.




\subsection{Surface topological orders and symmetry}

A 2+1D topologically ordered surface in the fermionic setting is described by a spin-TQFT, or, algebraically, a unitary pre-modular category ${\cal F}$ with a single transparent particle $f$, the local fermion.  Each such ${\cal F}$ has a modular completion, i.e. a larger unitary modular tensor category 

\begin{align}
{\cal B} = {\cal B}_0 \oplus {\cal B}_1
\end{align}
where ${\cal B}_0 = {\cal F}$.  Here the particles in ${\cal B}_1$ are the fermion parity $\pi$-fluxes, which all have non-trivial braiding $-1$ with the fundamental fermion $f$.

Now, a symmetry action of $G$ on such a unitary pre-modular category ${\cal F}$ is described, at the coarsest level, as a map from $G$ to the braided auto-equivalence group of ${\cal F}$.   A braided auto-equivalence $\phi$ of ${\cal F}$ is roughly just a permutation of the anyons in ${\cal F}$, $a \to \phi(a)$, that is compatible with the braiding and fusion structure. Formally, it is defined by also specifying unitary linear maps  on all of the fusion spaces,
\beq \phi  |a, b; c, \mu \rangle = \sum_{\nu} (\phi^{a,b}_c)_{\mu \nu} |\phi(a), \phi(b); \phi(c), \nu\rangle\eeq
where ${(\phi^{a, b}_c)}_{\mu \nu}$ is a unitary matrix. 
The braided auto-equivalence is only defined up to natural isomorphisms $\Upsilon$, such that $\phi \sim \Upsilon \circ \phi$,
\beq \Upsilon:\,\, |a, b; c\rangle \to \frac{\gamma(a) \gamma(b)}{\gamma(c)} |a, b; c\rangle\eeq
where $\gamma$ are $U(1)$ phases.
Thus, to each element $g\in G$ we assign a braided auto-equivalence $\phi_g$. We will alternatingly use the notation $\phi_g(a) \equiv \,^g\!a \equiv g \cdot a$. The group-law must be satisfied up to a natural isomorphism:
\beq \kappa_{g,h} \circ \phi_g \circ \phi_h = \phi_{gh} \label{eq:kappadef}\eeq
where 
\beq \kappa_{g,h}(a,b;c) = \frac{\beta_a(g,h) \beta_b(g,h)}{\beta_c(g,h)} \label{eq:kappabeta} \eeq
In the remainder of this section we will for simplicity only focus on the situation $\kappa_{g,h}=1$ for all $g,h$, in which case $\beta_a(g,h)$ can all be taken to be $1$ as well.  This is sufficient to describe all the examples we present, and we believe that the more general situation with non-trivial $\kappa_{g,h}$ can be handled with a slight extension of our formalism, see in particular section \ref{sec:H3gen}. For more information on braided auto-equivalences see Refs. \onlinecite{Maissam2014, ENO, Tarantino2015}.

In the `beyond cohomology' situation $[\sigma] \neq 0$, we conjecture that, although $G$ acts on ${\cal F}$ by braided auto-equivalences, these cannot be extended to braided auto-equivalences of its modular completion ${\cal B}$ in a way consistent with the group law.  Conversely, in the super-cohomology situation $[\sigma]=0$ we believe that the action by braided auto-equivalences can always be extended to the modular completion ${\cal B}$.  Let us now focus exclusively on this super-cohomology situation.

Given a valid action of $G$ by braided auto-equivalences, we can define another piece of data that characterizes the action of symmetry on the anyons, namely the symmetry fractionalization data.  In the case when $G$ does not permute the anyons at all - i.e. trivial braided-auto-equivalence action - such fractionalization data amounts to an assignment of a projective representation of $G$ to each anyon $a$, 
\beq U_a(g) U_a(h) = \omega_{a}(g,h) U_{a}(gh) \label{eq:Ughgh}\eeq
where $\omega_a(g,h)$ are $U(1)$ phase factors satisfying the usual co-cycle condition $d \omega_a(g, h, k) = 1$.  In the case of anyon-permuting symmetries, we can argue, using physical considerations similar to those of Ref. \onlinecite{Maissam2014}, that at the surface of a fermionic SPT the fractionalization data should still be encoded in a set of phases $\omega_a(g,h)$ satisfying now the twisted co-cycle condition:
\begin{align} \label{omega_eq}
1 = \omega_{g^{-1}\cdot a} (h,k) \omega_a(gh,k)^{-1} \omega_a(g,hk) \omega_a(g,h)^{-1}
\end{align}
as well as
\begin{align}
\omega_a(g,h)\omega_b(g,h) = \omega_c(g,h)
\end{align}
for any $c$ in the fusion product of $a$ and $b$. This is analogous to Eq. 164 of Ref. \onlinecite{Maissam2014}, where we have made use of our assumption $\kappa_{g,h}=1$ to simplify the left hand side. We will make an extra assumption that $\omega_f(g,h) = 1$  - indeed, the $f$ particle is a local excitation, so it does not transform projectively under the symmetry.

As in the bosonic case, the next step is to argue that we can represent this fractionalization data equally well as a collection of Abelian anyons $\tilde \omega(g,h)$ satisfying
\begin{align}
\omega_a(g,h) = M^*_{\tilde\omega(g,h),a}, \label{eq:omegaM}
\end{align}
with {$M^*_{ab} = \frac{S_{ab} S_{11}}{S_{1a} S_{1b}}$} the monodromy for the exchange of $a$ and $b$. When one of $a$ and $b$ is an Abelian anyon, $M^*_{ab}$ is a pure phase - the full-braid exchange statistics. We refer the reader to appendix \ref{anyon_proof} for a proof that  $\tilde{\omega} \in {\cal F}$ satisfying (\ref{eq:omegaM}) always exists.  $\tilde\omega(g,h)$ is not unique: rather, it is only unique up to fusion with the transparent fermion $f$.  Also, from Eq. \ref{omega_eq} we see that the anyon
\begin{align}
&\rho(g,h,k) \equiv d \tilde\omega(g,h,k) \nonumber\\
&= (g \cdot \tilde\omega(h,k)) \times \tilde\omega^{-1}(gh,k) \times \tilde\omega(g,hk) \times \tilde\omega^{-1}(g,h) \label{tomega_eq}
\end{align}
must have trivial braiding with all other anyons, i.e.

\begin{align}
\rho(g,h,k) \in \{1,f \}.
\end{align}
Now, because $\tilde\omega(g,h)$ was well defined only up to fusion with $f$, we see from Eq. \ref{tomega_eq} that $\rho$ is also well defined only up to the differential of a $2$-co-chain valued in $\{1,f\}$.  Furthermore, it is clear that $d\rho(g,h,k,j)=1$.  Thus, we have a well-defined cohomology class $[\rho] \in H^3(G,\Z_2)$.  

The crucial point is that $[\rho]$ might be non-zero.  Indeed, despite the fact that $\rho(g,h,k)=d \omega(g,h,k)$, it might be impossible to write $\rho$ as the differential of something $\Z_2$-valued.  In this case ($[\rho] \neq 1$), one cannot extend the fractionalization class to a valid fractionalization class in the modular completion ${\cal B}$ of ${\cal F}$, since such an extension is precisely a solution to
\begin{equation}
d \tilde \omega' (g,h,k) = 1
\end{equation}
where $\omega'(g,h)$ differs from $\omega(g,h)$ at most by a fermion $f$.  Here we have assumed for simplicity that $\kappa_{g,h}=1$ in the modular completion ${\cal B}$; again, this is true in the examples we consider, and our argument can presumably be generalized to the case of non-trivial $\kappa_{g,h}$ in the modular completion. 

Thus, a non-trivial $[\rho] \in H^3(G, \Z_2)$ signals an anomaly and shows that the bulk fermionic SPT order is non-trivial. The natural conjecture is that the bulk SPT order is also described by the same cohomology class $[\rho]$. Although we do not prove this conjecture in general, in the next section we will compute $[\rho]$ for the STO derived in the previous section for $G = \Z_2 \times \Z_4$ and show that it indeed matches the bulk SPT order.

We note that a non-trivial $[\rho]$ is only possible if the symmetry permutes the anyons. Indeed, we can always write the group of Abelian anyons 
$A$, as $A  = D \times \{1,f\}$, where $D$ is a subgroup of $A$. We can make a gauge choice where $\tilde{\omega}(g,h) \in D$ for all $g$, $h$. If the symmetry does not permute anyons then $\rho = d \tilde{\omega}(g,h,k) \in D$, which implies $\rho = 1$.

\section{Example: surface topological order for  $\Z_2 \times \Z_4 \times \Z_2^f$ SPT.} \label{sec:H3comp}

We now apply the formalism of the previous section to the conjectured STO of the $\Z_2 \times \Z_4 \times \Z_2^f$ SPT discussed in sections \ref{sec:STO}, \ref{sec:stack}. The STO is a $\Z_4$  gauge theory times a trivial fermionic theory $\{1,f\}$, with particles $m^j e^k f^\mu$, $j,k=0,\ldots,3$, $\mu=0,1$.  We will refer to this fermionic topological order as ${\cal F}$.  As discussed in the previous section, since we are at the surface of a fermionic SPT, we can specify the symmetry action at the level of braided auto-equivalence $\phi$ and fractionalization class $[\omega] \in H^2_\phi(G, A/\{1,f\})$, with $A$  - the group of Abelian anyons.  We now describe these in detail.

\subsection{Braided auto-equivalence action $\phi$}
\label{sec:phiZ2Z4}
For a group element $\bfg$ we will use the notation $\bfg = (\bfg_1,\bfg_2)$, $\bfg_1 \in \Z_2$, $\bfg_2 \in \Z_4$.  Similarly, we will represent an anyon $a$ as $a=[a_m,a_e,a_f]$:
\begin{align}
a = m^{a_m} e^{a_e} f^{a_f}
\end{align}
A group element $\bfg$ induces the following permutation action:

\begin{align} \label{perm_def}
a &\rightarrow ^{\bfg}\!\!a  \nonumber \\
[a_m,a_e,a_f] &\rightarrow [a_m, a_e+2 \bfg_2 a_m, a_f+\bfg_2 a_m]
\end{align}
Note that the permutation action of the $\Z_2$ generator is trivial.

At the level of splitting spaces $V^{a,b}$ we define
\begin{align}
\phi_{\bf g}: V^{a,b} \rightarrow V^{^\bfg a, ^\bfg b}
\end{align}
to act by the phase factor
\begin{align}
\phi^{a,b}_\bfg = (-1)^{\bfg_2 a_m b_f} (-1)^{{\left \lfloor{\frac{\bfg_2}{2}}\right \rfloor} a_m b_m} \label{eq:phig}
\end{align}
where $\lfloor x \rfloor$ denotes the largest integer smaller or equal than $x$. To see that these are valid braided auto-equivalences, we really only have to check the case ${\bf g}_2=1$, as ${\bf g}_2=2,3$ arise from taking powers of the case ${\bf g}_2 =1$.  In particular, the natural equivalences $\kappa_{{\bf g},{\bf h}}$ in Eq.~(\ref{eq:kappadef}) are all identically equal to $1$.

So we just have to check that $\phi_\bfg$ with $\bfg_2 = 1$ is a valid
braided auto-equivalence, i.e. that it commutes with all of the $F$ and $R$ symbols. The $F$ symbols are all trivial in this theory, and therefore commute with $\phi_{\bfg}$ because  $\phi^{a,b}_\bfg$ is a multiplicative bilinear form on ${\cal F} = \Z_4 \times \Z_4 \times \Z_2$.  Checking that $\phi_\bfg$ commutes with the $R$ symbols requires a computation.  First, we have
\begin{align}
R^{a,b} = \exp 2\pi i \left(\frac{a_m b_e}{4} + \frac{a_f b_f}{2} \right)
\end{align}
and therefore
\begin{align}
R^{^\bfg a,^\bfg b} = \exp 2\pi i \left(\frac{a_m(b_e+2b_m)}{4} + \frac{(a_f+a_m)(b_f+b_m)}{2} \right).
\end{align}
$\phi_\bfg$ commuting with the $R$ symbols means

\begin{align}
\phi^{b,a}_\bfg R^{a,b}= R^{^\bfg a,^\bfg b} \phi^{a,b}_\bfg
\end{align}
which reduces to
\begin{align}
\phi^{b,a} = (-1)^{a_m b_f + a_f b_m}\phi^{a,b}
\end{align}
which is indeed solved by $\phi$ in Eq.~(\ref{eq:phig}) with $\bfg_2 = 1$.

It is easy to show that the braided auto-equivalence action extends to a modular completion of this fermionic theory.  Indeed, one  modular completion of the trivial fermionic theory $\{1,f\}$ is just the toric code, so for the modular completion of ${\cal F}$ we can just take the theory ${\cal B}$ equal to a $\Z_4$ gauge theory times a $\Z_2$ gauge theory.  

For quasiparticles $b$ of ${\cal B}$ we will use the notation $b=[b_m,b_e,b_x,b_y]$, $b_m,b_e=0,\ldots, 3$, $b_x,b_y = 0,1$.  Here $[0,0,1,1]$ is the fundamental fermion $f$, and odd $b_x+b_y$ corresponds to the $\pi$-flux sector.  Again, the braided auto-equivalence action of $\bfg$ depends only on $\bfg_2$:
\begin{align}
b &\rightarrow ^{\bfg}\!\!b \nonumber \\
[b_m,b_e,b_x,b_y] &\rightarrow \nonumber \\ [b_m, b_e+2 \bfg_2(b_m+& b_x+b_y), b_x+\bfg_2 b_m, b_y+\bfg_2 b_m]
\end{align}

At the level of splitting spaces $V^{a,b}$ we define
\begin{align}
\phi_\bfg: V^{a,b} \rightarrow V^{^\bfg a, ^\bfg b}
\end{align}
to act by the phase factor
\begin{align}
\phi^{a,b}_\bfg = (-1)^{\bfg_2 a_m b_x} (-1)^{{\left \lfloor{\frac{\bfg_2}{2}}\right \rfloor} a_m b_m}
\end{align}
Just as before, one can explicitly check that this commutes with all the $F$ and $R$ symbols, and that all of the $\kappa_{\bfg,\bfh}$ natural isomorphisms are trivial.
 
\subsection{Fractionalization class $[\omega]$}
\label{sec:fracZ2Z4}

In the previous subsection we specified the action of symmetry $\Z_2 \times \Z_4$ by braided auto-equivalences, and found that all of the $\kappa_{\bfg,\bfh} = 1$, both in the fermionic theory ${\cal F}$ and in its modular extension ${\cal B}$.  Now we want to specify the symmetry fractionalization.  This amounts to specifying a function $\tilde\omega(g,h)$ from $G\times G$ to the anyons with the property that $d \tilde\omega(g,h,k) \in \{1,f \}$ for all $g,h,k$.  



Define:

\begin{align} \label{tomega_def}
\tilde\omega(\bfg,\bfh) = [ [\bfg_1 \bfh_1], \bfg_2 [\bfh_1],0] 
\end{align}

Here and below we let $[x] = 0$ if $x$ is even and $[x] = 1$ if $x$ is odd. Let us calculate $d\tilde\omega(\bfg,\bfh,\bfk)$, using Eq. \ref{tomega_eq}.

\begin{align}
\bfg \cdot \tilde\omega(\bfh,\bfk) &= [[\bfh_1 \bfk_1], \bfh_2 [\bfk_1] + 2 \bfg_2 \bfh_1 \bfk_1, \bfg_2 \bfh_1 \bfk_1] \\
\tilde\omega(\bfg \bfh,\bfk) &= [[(\bfg_1+\bfh_1)\bfk_1],(\bfg_2+\bfh_2)[\bfk_1],0] \\
\tilde\omega(\bfg,\bfh \bfk) &= [[\bfg_1(\bfh_1+\bfk_1)], \bfg_2 [\bfh_1+\bfk_1], 0] \\
\tilde\omega(\bfg,\bfh) &= [[\bfg_1 \bfh_1], \bfg_2 [\bfh_1],0]
\end{align}
A short calculation, utilizing the identity $[x] + [y] - [x+y] = 2 [x] [y]$, gives:

\begin{align} \label{rhodef}
\rho&(\bfg,\bfh,\bfk) = d \tilde \omega(\bfg,\bfh,\bfk)  \\ &= [0,0, \bfg_2 \bfh_1 \bfk_1]
\end{align}
i.e. $\rho(\bfg,\bfh,\bfk)$ is valued in $\{1,f\}$.  It is easy to see that $[\rho] \in H^3(\Z_2 \times \Z_4, \Z_2)$ is non-zero.  Indeed, it is the product of the generator of $H^1(\Z_4, \Z_2)$ with the generator of $H^2(\Z_2, \Z_2)$ in the cohomology ring, which is just a tensor product of the individual cohomology rings of $\Z_4$ and $\Z_2$.  This is the same $H^3$ cohomology class that describes the bulk SPT order for $G=\Z_2 \times \Z_4$ in Ref. \onlinecite{CTW}.

\subsection{Relation to the topological order constructed for the crystalline $\Z_2 \times \Z \times \Z_2^f$ SPT}

Let us compare the topological order we formally described in sections \ref{sec:phiZ2Z4},\ref{sec:fracZ2Z4} to one constructed on the surface of the crystalline $\Z_2 \times \Z \times \Z_2^f$ SPT in section \ref{sec:stack}. First, it is believed that for SETs with translation symmetry, one may formally treat translations as an internal symmetry $\Z$ for the purpose of specifying the symmetry data.\cite{Cheng_Bonderson}  Observe that the  $\phi_{\bfg}$ (\ref{perm_def}), (\ref{eq:phig}) and $\omega_{\bfg}$ (\ref{tomega_def}) are still consistent if we replace $\Z_4 \to \Z$. Furthermore, $\rho(\bfg,\bfh,\bfk) = d \tilde{\omega}(\bfg,\bfh,\bfk)$ is still a non-trivial co-cycle in $H^3(\Z_2 \times \Z, \Z_2)$. Now, this is actually the same STO as constructed in section \ref{sec:stack}.  Indeed, it has the same quasiparticles and the same permutation action of the symmetry. It remains to check that it has the same fractionalization data, table \ref{proj}.

Recall, for anyon $b$ which is not permuted by symmetry elements $\bfg$ and $\bfh$, the projective action of the symmetry on $b$ is given by  Eqs.~(\ref{eq:Ughgh}), (\ref{eq:omegaM}). 
From Eq. \ref{tomega_def} we have $\tomega(\bfg, \bfh) = (\bfg_1 \bfh_1, \bfg_2 [\bfh_1],0) = m^{[\bfg_1 \bfh_1]} e^{\bfg_2 [\bfh_1]}$, with $\bfg_1, \bfh_1 \in \Z_2$ and $\bfg_2, \bfh_2 \in \Z$. In particular, $\tomega([1,0], [1,0]) = m$, so $U^2_e(\mfg_1) =  \omega_e([1,0],[1,0]) = M^*_{e,m} = e^{2 \pi i/4}$ and $U^2_m(\mfg_1) = \omega_m([1,0], [1,0]) = M^*_{m,m}= 1$. This agrees with $e$ carrying charge $\frac14$ under $\Z_2$ and $m$ carrying no fractional charge. Next, we proceed to the ``commutator" of $\Z$ and $\Z_2$. We have $U^a_\bfg U^a_\bfh (U^a_\bfg)^{-1} (U^a_\bfh)^{-1} = \omega_a(\bfg,\bfh) \omega^*_a(\bfh,\bfg) = M^*_{a, \tomega(\bfg,\bfh) \times \tomega^{-1}(\bfh,\bfg)}$. Now, $\tomega([1,0],[0,1]) = 1$, $\tomega([0,1],[1,0]) = e$, $\tomega([1,0],[0,1]) \times \tomega^{-1}([0,1],[1,0]) = e^3$. Therefore, the commutator of $\mfg_1 \sim [1,0]$ and $T_a \sim [0,1]$ on $e$ is $[\mfg_1, T_a] = M^*_{e,e^3} = 1$ - trivial, and on $m^2$, $[\mfg_1, T_a] = M^*_{m^2, e^3} = -1$ - in agreement with table \ref{proj}. Finally, $\tomega([1,0],[0,2]) = 1$, $\tomega([0,2],[1,0]) = e^2$, $\tomega([1,0],[0,2]) \times \tomega^{-1}([0,2],[1,0]) = e^2$, so the commutator of $\mfg_1$ and $T^2_a$ on $m$ is $[\mfg_1, T^2_a] = M^*_{m, e^2} =-1$, again in agreement with table \ref{proj}. 

\subsection{$n = 2$ STO}
We now perform one more check on our STO for the $\Z_2 \times \Z_4 \times \Z^f_2$ SPT. Two copies, $n  = 2$, of this fermion SPT give rise to a purely bosonic SPT, as can be checked e.g. from the three-loop braiding data in Eq.~(\ref{eq:3loop}). Indeed, by doubling the phases in (\ref{eq:3loop}), we see that the $n =2$ state will be characterized by
\beq 2 \theta_{\mfg_1; \mfg_2} = \pi \eeq
with all the other braiding invariants being trivial. But this is precisely the braiding data of a bosonic SPT with $G = \Z_2 \times \Z_4$ symmetry corresponding to a co-cycle $\nu \in H^4(G, U(1))$ with 
\beq \nu(\bfg, \bfh, \bfk, \bfl) = \exp(\pi i \bfg_2 \bfh_1 \bfk_1 \bfl_1), \label{eq:nudouble}\eeq
see Ref.~\onlinecite{WL2015}. Let us confirm that this is consistent with the STO we've constructed for the $n =1$ phase.

If we stack two copies of the $n  =1$ STO, we get a theory with $\Z^{(1)}_4 \times \Z^{(2)}_4 \times \{1, f\}$ topological order with the first two factors generated by ($e_1$, $m_1$) and ($e_2$, $m_2$) respectively. 
The symmetry transformations are inherited from each STO copy, $\Z_4:\, m_1 \to m_1 e^2_1 f,\,m_2 \to m_2 e^2_2 f$. 
Likewise, the projective factors $\tilde{\omega}_a(\bfg,\bfh)$ are inherited from each copy, so that the class $\tilde{\omega}$ is
\beq \tilde{\omega}(\bfg,\bfh) =  (m_1 m_2)^{[\bfg_1 \bfh_1]} (e_1 e_2)^{\bfg_2 [\bfh_1]} \label{eq:tomega2}\eeq 
It is easy to check that now $d \tilde{\omega} = 0$, i.e. there is no $H^3$ anomaly. This suggests that we are dealing with the surface of a bosonic SPT. We would like to find out if this SPT is trivial or not. To do so, it is convenient to first simplify the STO by driving an anyon condensation transition. Let us condense the anyon $e_1 e^3_2$. After eliminating the confined anyons and performing anyon identification, we are left with a $\Z_4 \times \{1, f\}$ topological order generated by $\tilde{e} = e_1  \sim e_2 $, $\tilde{m} = m_1 m_2$ and $f$. Note that the condensing anyon $e_1 e^3_2$ is not permuted by the symmetry and carries a trivial fractionalization $\omega_{e_1 e^3_2}(\bfg, \bfh)  = 1$. Thus, the condensation does not break the symmetry. After the condensation none of the anyons are permuted by the symmetry since now $\Z_4:\,\tilde{m} \to m_1 m_2 e^2_1 e^2_2 \sim \tilde{m}$. The fractionalization class $\tilde{\omega}$ (\ref{eq:tomega2}) now becomes
\beq \tilde{\omega}(\bfg, \bfh) = \tilde{m}^{[g_1 h_1]} \tilde{e}^{2g_2  h_1} \eeq
i.e. $\tilde{e}$ carries charge $\frac14$ under $\Z_2$, and $\mfg_1$, $\mfg_2$ anticommute on $\tilde{m}$. Clearly, the fermion $f$ now plays no role in the resulting topological order and symmetry action, so this is an STO of a bosonic SPT. Since the anyons are not permuted by the symmetry, we can use the result of Ref.~\onlinecite{ProjS} to compute the co-cycle $\nu \in H^4(G, U(1))$ characterizing the bulk bosonic SPT:
\bea &&\nu(\bfg, \bfh, \bfk, \bfl) = R_{\omega(\bfk, \bfl), \omega(\bfg, \bfh)} \times \nonumber \\
&&F_{\omega(\bfh, \bfk), \omega(\bfg, \bfh \bfk), \omega(\bfg \bfh \bfk, \bfl)} F^{-1}_{\omega(\bfh, \bfk), \omega(\bfh\bfk, \bfl), \omega(\bfg, \bfh \bfk \bfl)} \nonumber \\
&&F_{\omega(\bfg, \bfh), \omega(\bfk,  \bfl), \omega(\bfg \bfh, \bfk \bfl)} F^{-1}_{\omega(\bfg, \bfh), \omega(\bfg \bfh, \bfk), \omega(\bfg \bfh \bfk, \bfl)} \nonumber \\
&&F_{\omega(\bfk, \bfl), \omega(\bfh, \bfk \bfl), \omega(\bfg, \bfh \bfk \bfl)} F^{-1}_{\omega(\bfk, \bfl), \omega(\bfg, \bfh), \omega(\bfg \bfh, \bfk \bfl)} \nonumber \eea
Now, in the $\Z_4$ gauge theory all the $F$ symbols are trivial, so using $R_{a,b} = \exp(2 \pi i a_m b_e/4)$, we, indeed, reproduce Eq.~(\ref{eq:nudouble}). 

\section{General $H^3$ obstruction}
\label{sec:H3gen}

In this section we describe a general formalism that unites the $H^3(G,\Z_2)$ anomaly introduced in this paper with other $H^3$ anomalies introduced in other contexts.  Let ${\cal A}' \subset {\cal A}$ be braided fusion categories and $G$ a finite unitary symmetry group.  Suppose for now that ${\cal A}$ is modular, although later we will generalize beyond this.  Suppose we have an action of $G$ by braided auto-equivalences on ${\cal A}$, and that ${\cal A}'$ is preserved under this action, so that there is an induced action by braided auto-equivalences on ${\cal A}'$.  Let $A$ and $A'$ denote the Abelian groups of Abelian anyons in ${\cal A}$ and ${\cal A}'$ respectively; in particular $A' \subset A$.  Suppose that there is no $H^3(G,A')$ obstruction to localizing the symmetry on ${\cal A}'$ and defining a symmetry fractionalization class for those anyons.  One can then ask: what is the obstruction to extending this fractionalization class to all of ${\cal A}$?  We will see that in general it is valued in $H^3(G,T)$, where $T \subset A$ is the subgroup of $A$ consisting of Abelian anyons that have trivial braiding with ${\cal A}'$ (there is an additional mild technical assumption that we spell out below).  

Before we give the argument, note that several different examples given in the literature are special cases of this obstruction.

{\bf(1)} If we choose ${\cal A}' = \{1\}$ to be the trivial category, then $T=A$, and the obstruction is valued in $H^3(G,A)$; this is just the original $H^3(G,A)$ obstruction of Ref. \onlinecite{ENO}.

{\bf(2)} If ${\cal A}'$ is a fermionic theory and ${\cal A}$ its modular completion, then $T = \{1,f\} = \Z_2$ and we have the $H^3(G,\Z_2)$ obstruction given in this paper.

{\bf(3)} Let ${\cal A}$ be a $\Z_3$ gauge theory with non-trivial Dijkgraaf-Witten twist, so that its anyons form the group $\Z_9$ under fusion.  It is natural to label the generator of this $\Z_9$ as $m$.  Then let ${\cal A}'$ be the $\Z_3$ subgroup of gauge charges $\{ 1,e,e^2 \} = \{ 1,m^3,m^6 \}$, where we have defined $e=m^3$.  Then Ref. \onlinecite{KapustinThorngrenZ3} shows that certain extensions of the gauge group $\Z_3$ by the symmetry group $G=\Z_3 \times \Z_3$ are inconsistent with the twist.  But this extension data  just specifies a fractionalization class of $G$ on ${\cal A}'$, for a trivial braided autoequivalence action on ${\cal A}$ and ${\cal A}'$.  Hence according to our general result this inconsistency is encoded in $H^3(G,T) = H^3(G,\Z_3)$, since $T={\cal A}' = \Z_3$.


{Returning to the general formalism,} to derive our obstruction, start with the natural isomorphisms $\kappa_{f,g}$ defined in Eq.~(\ref{eq:kappadef}). 
These restrict to natural isomorphisms on ${\cal A}'$. The action of $\kappa_{f,g}$ on splitting spaces can be decomposed into factors $\beta_a(f,g)$, see Eq.~(\ref{eq:kappabeta}), 
from which one constructs

\begin{align} 
\Omega_a(g,h,k) = \frac{\beta_{g^{-1} \cdot a} (h,k) \beta_a(g,hk)}{\beta_a(gh,k) \beta_a(g,h)}
\end{align}
and finds that

\begin{align}
\Omega_a(g,h,k) = M^*_{a, O(g,h,k)}
\end{align}
where $O(g,h,k) \in {\cal A}$ is an Abelian anyon.

Now suppose that, when restricted to the anyons in ${\cal A}'$, the symmetry can be localized.  Precisely stated, this assumption means that one can find phase factors $\nu_{a'}(g,h)$, $a' \in {\cal A}'$, satisfying $\nu_{a'}(g,h) \nu_{b'}(g,h) = \nu_{c'}(g,h)$ whenever $N^{a'b'}_{c'} \neq 0$, such that 

\begin{align}
\Omega_{a'}(g,h,k) = \frac{\nu_{g^{-1} \cdot a'} (h,k) \nu_{a'}(g,hk)}{\nu_{a'}(gh,k) \nu_{a'}(g,h)}
\end{align}
for all $a' \in {\cal A}'$.  The question we want to ask is: given $\{\nu_{a'}(g,h) \}_{a' \in {\cal A}'}$, what is the obstruction to extending them to a full set of $\{\nu_{a}(g,h) \}_{a \in {\cal A}}$ satisfying $\nu_{a}(g,h) \nu_{b}(g,h) = \nu_{c}(g,h)$ whenever $N^{ab}_c \neq 0$?  In essence, given a choice of symmetry localization on ${\cal A}'$, what is the obstruction to extending it to all of ${\cal A}$?

To answer this question, we first have to make a technical assumption: we assume there exist Abelian anyons $o(g,h) \in A$ such that

\begin{align} \nu_{a'}(g,h) = M^*_{a', o(g,h)} \end{align}
for all $a' \in {\cal A}'$.  Certainly this assumption holds for Abelian theories ${\cal A}$.  Also, there definitely exist non-Abelian theories where this assumption fails for some such multiplicative functions $\nu_{a'}$ (e.g. $\{1,f \} \subset \{1,f,\sigma\}$).  
In any case, assuming the existence of $o(g,h)$, it is clear that it is well defined up to fusion with anyons in $T$, where $T$ is the subset of Abelian anyons $A$ which have trivial braiding with ${\cal A}'$.

Define now

\begin{align}
\tilde{O}&(g,h,k) \nonumber  \\
 &= O(g,h,k) (g \cdot o(h,k))^{-1} o(g,hk)^{-1} o(gh,k) o(g,h)
\end{align}
Then $\tilde{O}(g,h,k)$ has the key property that its braiding with any anyon in ${\cal A}'$ is trivial.  Hence $\tilde{O}(g,h,k) \in T$.  Finally, note that $\tilde{O}(g,h,k)$ is clearly a co-cycle, and since $o(g,h)$ is ambiguous up to fusion with anyons in $T$, $\tilde{O}(g,h,k)$ is only well defined up to the co-boundary of a $T$-valued $2$-co-chain.  Hence $\tilde{O}(g,h,k)$ determines a cohomology class

\begin{align}
[\tilde{O}] \in H^3(G,T)
\end{align}

\subsection{Special case}
In the special case where the braided auto-equivalence action on ${\cal A}$ has no $H^3(G,A)$ obstruction, we can make the above discussion slightly more concrete.  In this case we can find decomposition factors $\beta_a(g,h)$ for which $\Omega_a(g,h,k)$ vanishes identically.  Then a choice of symmetry localization on ${\cal A}'$ is precisely a choice of phase factors $\omega_{a'}(g,h)$, $a' \in {\cal A}'$, satisfying $\omega_{a'}(g,h) \omega_{b'}(g,h) = \omega_{c'}(g,h)$ whenever $N^{a'b'}_{c'} \neq 0$, such that 

\begin{align} \label{omega_special_case}
\frac{\omega_{g^{-1} \cdot a'} (h,k) \omega_{a'}(g,hk)}{\omega_{a'}(gh,k) \omega_{a'}(g,h)} = 1
\end{align}
for all $a'$.  Again, we want to ask: what is the obstruction to extending $\nu_{a'}$ to all of ${\cal A}$?  To answer this, we again make the technical assumption that there exists $o(g,h) \in A$, well defined up to anyons in $T$, such that $\omega_{a'}(g,h) = M^*_{a', o(g,h)}$ for all $a' \in {\cal A}'$.  Viewing $o(g,h)$ as a well defined element of $A/T$, Eq. \ref{omega_special_case} shows that it satisfies the co-cycle equation $do = 1$ and hence determines a cohomology class $[o] \in H^2(G,A/T)$.  Then the class $[\tilde{O}] \in H^3(G,T)$ can be obtained from $[o]$ by using the map

\begin{align} H^2(G,A/T) \rightarrow H^3(G,T), \end{align}
which is the co-boundary map induced from the exact coefficient sequence

\begin{align}
1 \rightarrow T \rightarrow A \rightarrow A/T \rightarrow 1
\end{align}

\section{Discussion}
\label{sec:disc}

We conclude by pointing out some open questions.

1. In this paper, we have defined a new anomaly class $[\rho] \in H^3(G, \Z_2)$ for 2+1D fermion topological orders. We conjecture that topological orders existing on the surface of a 3+1D fermion super-cohomology SPT possess such an anomaly and that $[\rho]$ matches the bulk co-cycle describing the SPT. We leave the proof of this conjecture to future work. One possible direction is to link the surface anomaly to three-loop braiding characterizing the bulk SPT. For boson SPTs it is known how to do this in the case when the symmetry does not permute the anyons in the STO.\cite{WLL} However, $[\rho] \neq 0$ necessarily requires the symmetry to permute the anyons, making this approach potentially challenging.

2. It is currently not clear if a fermion topological order with any $[\rho] \in H^3(G, \Z_2)$ can be realized at the surface of some 3+1D fermion SPT  or if there are additional constraints that the STO must satisfy.  For instance, from the bulk classification we know that the bulk co-cycle $\rho$ must satisfy $\frac12 \rho \cup_1 \rho = d\nu$, for some $\nu \in C^4(G, \mathbb{R}/\Z)$. It is not obvious that the surface anomaly class $[\rho]$ we've defined necessarily has this property.

3. 
A  general question is whether every 3+1D fermion SPT with a finite symmetry group admits a topologically ordered symmetric surface. For SPTs of bosons, Ref.~\onlinecite{WWW} has answered this question in the affirmative by providing a systematic construction of exactly solvable models of STOs. The idea is to write the symmetry group $G$ as $G = H/K$, so that viewed as a $H$-SPT the bulk phase is trivial. Ref.~\onlinecite{WWW} then shows how to construct the STO as a $K$-gauge theory. It is interesting if this construction can be generalized to fermion SPTs. 

4. In this paper, we have only considered fermion SPTs in the super-cohomology classification. We expect that for such SPTs the topological symmetry action in the STO can be extended to fermion parity fluxes, but the symmetry fractionalization cannot. For the fermion SPTs outside super-cohomology we conjecture that the surface anomaly is more severe and the topological symmetry action cannot be extended to the fermion parity fluxes. It would be interesting to construct examples of such STOs. Since the beyond super-cohomology SPTs are characterized by $\sigma \in H^2(G, \Z_2)$, we expect there to be a surface anomaly indicator of this type.

\section{Acknowledgements}
We would like to thank Chenjie Wang and Michael Zaletel for useful discussions. We would also like to thank  the organizers and staff of the Simons Center for Geometry and Physics workshop  ``Mathematics of topological phases of matter" where this work was initiated.  AV was supported by a Simons Investigator Grant.  LF was supported by NSF DMR-1519579.

\begin{appendix}

\section{Crystalline topological insulator with $\Z_2 \times \Z \times U(1)$ symmetry and its STO}
\label{app:U1}

In this section, we  generalize the construction in section \ref{sec:stack} to the case when the system has a full $U(1)$ particle number-symmetry. The full symmetry group then becomes $\Z_2 \times \Z \times U(1)$ with  $\Z^f_2 \subset U(1)$. We find an interesting interplay between the $U(1)$ symmetry and the projective representation of $\Z_2 \times \Z$ carried by the anyon $m^2$.

We begin with the stack construction of the bulk. The $\nu =2$ SPT layers now have both right and left-movers charged under $U(1)$, while the quantum numbers under $\Z_2$ are unchanged:
\beq U(1)_\alpha: \nu_{R/L} \to \nu_{R/L} + \alpha \eeq

Next we proceed to the surface. For the strips of $\Z_4 \times \{1,f\}$ topological order, we choose $e$ to have $U(1)$ charge $\frac12$ and $m$ - charge $0$:
\beq U(1)_\alpha: \phi \to \phi + \frac{\alpha}{2}, \,\, \theta \to \theta\eeq
Note that with this assignment all local bosons have even integer charge under $U(1)$ as required. We leave the $\Z_2$ charge assignments as before, Eq.~(\ref{Z2edge}).

Let us focus on a single T-junction. Now, $\Phi = (\theta, \phi, \nu_R, \nu_L)$ carries $U(1)$ charge $(0,\frac12, 1, 1)$. So does $\Phi'$ in Eq.~(\ref{eq:Sdef}). Therefore, by going to the $\Phi'$ variables, we see that the $\Z_4 \times \{1, f\}$ topological order with the $\nu  =2$ $\Z_2 \times U(1)$ SPT on top is identical to one with no SPT on top. In particular, the perturbation (\ref{gapnup}) gaps out the extra modes $\nu'_R$, $\nu'_L$.

Finally, to construct the STO we, as before, stitch the strips of $\Z_4$ topological order with the gapping terms in Eq.~(\ref{gap3}). Note that these terms preserve the $U(1)$ symmetry.

How does the $U(1)$ symmetry in the bulk and on the surface interplay with $\Z_2 \times \Z$ symmetry? To answer this question, let us study the properties of the magnetic monopole in the $3+1$D bulk of this phase. Imagine we have a pair of magnetic monopoles in the $3+1$D bulk, each sitting between two consecutive $\nu = 2$ $xz$ planes and separated by a distance $d$. When the $xz$ planes are infinite, each plane between the two monopoles sees a total magnetic flux $2\pi$, while all other planes see magnetic flux $0$. Now, a property of the $\nu = 2$ $\Z_2 \times U(1)$ SPT is that a $2\pi$ flux carries a $\Z_2$ charge $1$.\footnote{Indeed, think of $\nu =2$ as two layers of IQH effect, one  with $\sigma_{xy} = 1$ and the other with $\sigma_{xy} = -1$. Only the electrons in the first layer are charged under $\Z_2$. A $2\pi$ flux induces charge $1$ in the first layer and charge $-1$ in the second layer. Since only the first layer is charged under $\Z_2$, this translates into a $\Z_2$ charge of $1$ (and $U(1)$ charge of $0$).} Therefore, the $\Z_2$ charge of the monopole-antimonopole configuration changes by $1$ as the distance $d$ is increased by $1$. We conclude that $\Z_2$ and the generator of translations $\Z$ anti-commute on the monopole. What is the STO analogue of this effect? When a monopole tunnels across the surface of the 3+1D SPT, it leaves the anyon $m^2$ on the surface.\footnote{Indeed, recall that the anyon $b$ created by the tunnelling event must satisfy $e^{2 \pi i q_a} = M_{ab}$ for all anyons $a$, where $q_a$ is the anyon charge. The anyons $m^2$ and $m^2 f$ are the only ones that satisfy this criterion. If we consider tunnelling of a neutral bulk monopole then the neutral anyon $m^2$ is left on the surface.} As we saw in section \ref{sec:stack}, the generators of $\Z_2$ and  $\Z$ anti-commute on $m^2$, just as they do on the bulk monopole.

\begin{figure}
\begin{center}
\includegraphics[width = 3.5in]{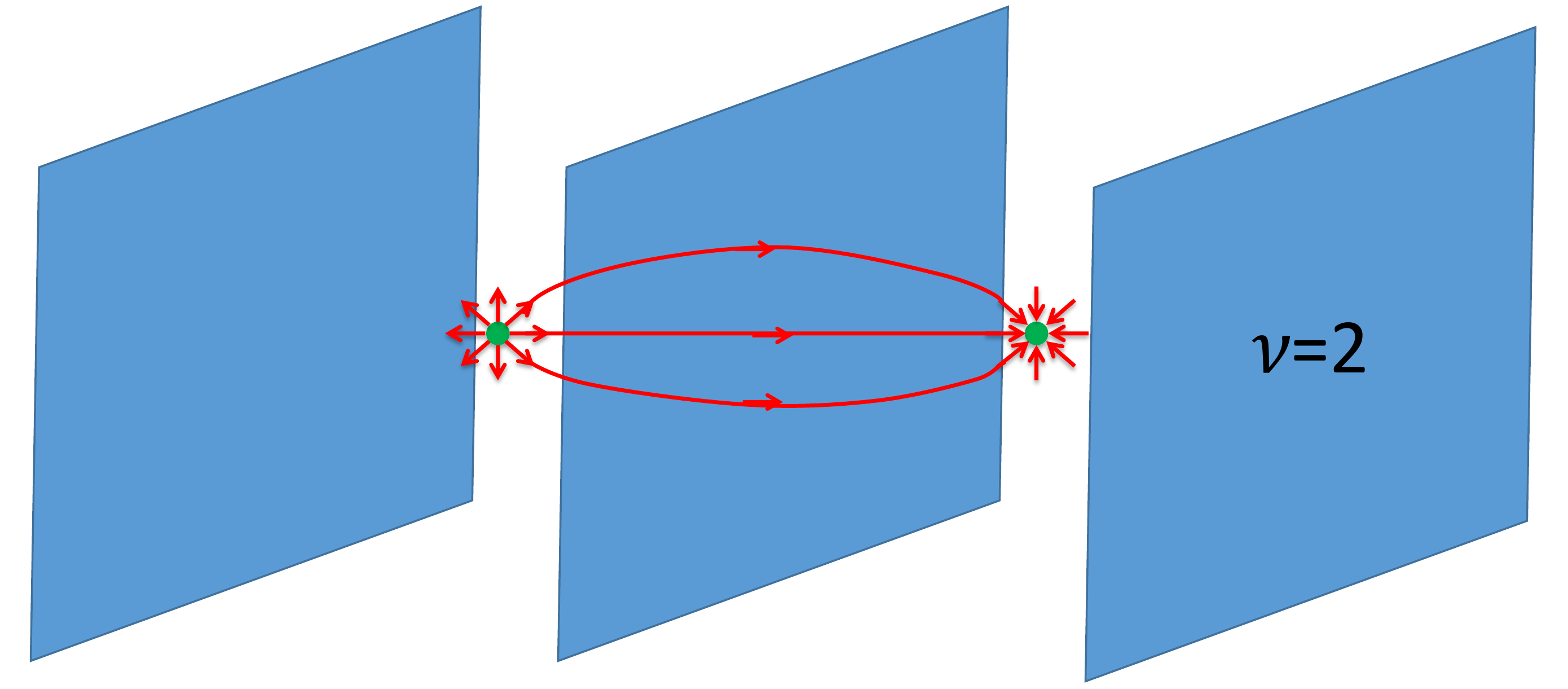}
\end{center}
\caption{$U(1)$ monopole-antimonopole pair in the bulk of the crystalline  SPT with $\Z_2 \times \Z \times U(1)$ symmetry. The $\Z_2$ charge carried by this configuration is equal to the number of $\nu =2$ $\Z_2 \times U(1)$ SPT layers between the monopoles, i.e. the generators of $\Z_2$ and $\Z$ anti-commute on the monopole. }
\end{figure}

\section{Proof of existence of ${\tilde{\omega}}(g,h)$ in a fermionic theory} \label{anyon_proof}

Suppose we have a set of phases $\omega_a \in U(1)$ defined for anyons $a \in {\cal F}$ and satisfying $\omega_a  \omega_b = \omega_c$ for any $a, b, c$ with $N^c_{ab} \neq 0$. Furthermore, assume $\omega_f = 1$. We claim that there exists an Abelian anyon $b \in {\cal F}$ with $\omega_{a} = M^*_{a b}$. Furthermore $b$ is unique up to fusion with $f$. The proof proceeds along essentially the same lines as in the bosonic case (see section II of Ref. \onlinecite{Maissam2014}). Indeed, let $\lambda_a  = d_a \omega_a$. Then $\lambda_a$ is a character of the fusion algebra of ${\cal F}$, i.e. it satisfies
\beq \lambda_a \lambda_b = \sum_c N^c_{ab} \lambda_c \eeq
Indeed, for $c$ contributing to the sum above $\lambda_c = d_c \omega_c = d_c \omega_a \omega_b$ and $\sum_c N^c_{ab} d_c = d_a d_b$. Now, the algebra should admit exactly $|{\cal F}|$ characters - the number of anyons in $|{\cal F}|$. By Verlinde formula, at least some of the characters can be obtained from the $S$-matrix of ${\cal F}$: an anyon $b \in {\cal F}$ generates a character, $\lambda^{(b)}_a = \frac{S_{ab}}{S_{1b}}$. In fact, this generates all the characters with $\lambda_f = 1$ (of which there are $|{\cal F}|/2$), i.e. the characters that we are interested in.

To see this, it is actually more convenient to work with an abstract theory where anyons $a$ and $a f$ of ${\cal F}$ are identified.\cite{WenFerm} This theory, which we call $\tilde{\cal F}$, has half the anyons of ${\cal F}$. Fusion in $\tilde{{\cal F}}$ is defined as $\tilde{N}^{[c]}_{[a],[b]} = N^c_{ab} + N^{cf}_{ab}$, and  does not depend on the representatives $a,b,c$ of equivalence classes $[a],[b],[c]$.  Fusion in $\tilde{{\cal F}}$ is associative, commutative and each element possesses a unique inverse. We can also form a modified $S$-matrix, $\tilde{S}_{[a],[b]} = \sqrt{2} S_{ab}$, which again does not depend on the choice of representatives. In fact, $\tilde{S}$ is a unitary matrix.\cite{WenFerm}  Any character of ${\cal F}$ with $\lambda_f = 1$ projects to a character of $\tilde{F}$. Thus, $\lambda^{(b)}_a = \frac{S_{ab}}{S_{1b}} = \frac{\tilde{S}_{ab}}{\tilde{S}_{1b}}$ are characters of $\tilde{F}$. Since $\tilde{S}$ is unitary, this provides all $|\tilde{{\cal F}}| = \frac{|{\cal F}|}{2}$ distinct characters of $\tilde{{\cal F}}$. Therefore, we conclude that our $\omega_a$ can be written as $\omega_{a} = \frac{S_{ab} }{d_a S_{1b}} = \frac{S_{ab} S_{11}}{S_{1a} S_{1b}} = M^*_{ab}$ for some $b \in {\cal F}$, which is unique up to $b \to b f$. 

It remains to show that if $\omega_a = M^*_{ab}$ is a pure phase for all $a \in {\cal F}$ then $b$ must be Abelian. As in the bosonic case,\cite{Maissam2014} this follows from unitarity of $\tilde{S}$. Indeed,
\bea 1 &=& \sum_{a \in \tilde{\cal F}} |\tilde{S}_{a b}|^2 = 2 \sum_{a \in \tilde{{\cal F}}} |S_{ab}|^2 = \sum_{a \in {\cal F}} |S_{ab}|^2 = \sum_{a \in {\cal F}} \left|\frac{d_a d_b M_{ab}}{{\cal D}}\right|^2 \nonumber \\
&=& d^2_b\eea
So $d_b  =1$.

\end{appendix}

\bibliographystyle{h-physrev}
\bibliography{fSTO}

\end{document}